\documentclass[onecolumn,apj,numberedappendix]{emulateapj}
\newcommand{\be}{\begin{equation}}
\newcommand{\ee}{\end{equation}}
\newcommand{\ba}{\begin{eqnarray}}
\newcommand{\ea}{\end{eqnarray}}
\def \d {{\rm d}}

\submitted{ApJ, accepted 26 July 2006}
\begin{document}

\title{Studying Turbulence using Doppler-broadened lines: Velocity 
Coordinate Spectrum}

\author{A. Lazarian}
\affil{Department of Astronomy, University of Wisconsin,
Madison, US}
\author{D. Pogosyan}
\affil{Physics Department, University of
Alberta, Edmonton, Canada}

\begin{abstract}
We discuss a new technique for studying astrophysical turbulence
that utilizes the statistics
of Doppler-broadened spectral lines. The technique relates
the power Velocity Coordinate Spectrum (VCS), i.e.
the spectrum of fluctuations measured along the velocity axis 
in Position-Position-Velocity (PPV) data cubes available from observations, 
to the underlying power spectra
of the velocity/density fluctuations. Unlike the standard spatial
spectra, that are function of angular wavenumber, the VCS is a
function of the velocity wave number $k_v\sim 1/v$.
We show that absorption affects the VCS to a higher degree
 for small $k_v$ and obtain the
criteria for disregarding the absorption effects for turbulence studies at
large $k_v$. 
We consider the retrieval of turbulence spectra from observations for high
and low spatial resolution observations and find that the VCS allows one
to study turbulence even when the emitting turbulent volume is not spatially
resolved. This opens interesting prospects for using the technique for 
extragalactic research.
We show that, while thermal broadening interferes with the turbulence
studies using the VCS, it is possible to
separate thermal and non-thermal contributions. This allows a new way of
determining the temperature of the interstellar gas using emission and
absorption spectral lines. 
\end{abstract}


\section{Introduction}

Astrophysical fluids are usually turbulent and the turbulence 
is magnetized. This ubiquitous turbulence determines the transport
of heat and cosmic rays in the interstellar medium (see 
Elmegreen \& Falgarone 1996, Stutzki 2001, Narayan \& Medvedev 2001,
Schlickeiser 2002, Cho et al. 2003, Lazarian 2006a) 
the intra-cluster medium
(Inogamov \& Sunyaev 2003, Sunyaev, Norman \& Bryan 2003),
processes of star formation (see McKee \& Tan 2002, 
Elmegreen 2002, Larson 2003, Ballesteros-Paredes et al. 2006), 
and interstellar chemistry (see Falgarone 
1999, Falgarone et al. 2006). 
An extended list of interstellar
processes governed by turbulence is given in Elmegreen \& Scalo (2004).

Using a statistical description is a nearly indispensable strategy when
dealing with turbulence. The big advantage of statistical techniques
is that they extract underlying regularities of the flow and reject
incidental details. Kolmogorov description of unmagnetized incompressible
turbulence is a statistical
one. For instance, it predicts that the difference in velocities at
different points in a turbulent fluid increases on average
with the separation between points as a cube root of the separation,
i.e. $|\delta v| \sim l^{1/3}$. In terms of the direction-averaged
energy spectrum this gives the famous Kolmogorov
scaling $E(k)\sim 4\pi k^2 P( k)\sim k^{5/3}$, where $P( k)$ 
is a {\it 3D} energy spectrum defined as the Fourier transform of the
correlation function of velocity fluctuations $\langle  
\delta v({\bf x})\delta v({\bf x}+{\bf r})\rangle$. In
this paper we use $\langle  ...\rangle$ to denote the ensemble 
averaging procedure\footnote{The relation between the ensemble averaging and 
more common spatial averaging is discussed in Monin \& Yaglom (1975). We address the related issues throughout the paper when we discuss the practical implementations of our statistical technique.}.

The velocity energy spectrum $E(k)dk$ characterizes how much
energy resides in the interval of scales $k, k+dk$. At large scales $l$
which correspond to small wave-numbers $k$ ( i.e. $l\sim 1/k$) one expects
to observe features reflecting energy injection. At small scales
one should see the scales corresponding to
sinks of energy. In general, the shape of the spectrum is
determined by a complex process of non-linear energy transfer and
dissipation. For Kolmogorov turbulence the spectrum 
over the inertial range, i.e. the range where neither  energy injection nor
energy dissipation are important, is
characterized by a single power law and is, therefore, self-similar.
Other types of turbulence, i.e. the turbulence of non-linear waves
or  the turbulence of shocks, are characterized by different power
laws and therefore can be distinguished from the Kolmogorov turbulence
of incompressible eddies.
Substantial advances in our understanding of the scaling of compressible
MHD turbulence (see reviews by Cho \& Lazarian (2005),
and references therein) allows us to provide a direct comparison of
theoretical expectations with observations.

Recovering the velocity spectra from observations
 is a challenging problem that
has been studied for more than half a century.
Indeed, the first measurements were obtained with
the velocity in the 50s 
(see von Hoerner 1951, Munch 1958). While the centroids were widely used
to study turbulence in molecular clouds (see 
Kleiner \& Dickman 1985, Dickman \& Kleiner 1985, Miesch \& Scalo 1995, 1999) 
the recent theoretical work and numerical testing
(Lazarian \& Esquivel 2003, Esquivel \& Lazarian 2005, Ossenkopf et al. 2006) 
show that velocity centroids can reliably recover
velocity spectra only for subsonic or mildly supersonic
turbulence.

If turbulence is supersonic, the recovering of its velocity spectrum is
possible with the Velocity Channel Analysis (VCA), introduced in LP00.
This technique has already been successfully used to study turbulence
(see Stanimirovic \& Lazarian 2001).
However, the VCA is just one way to use
the general description of 
fluctuations in the Position-Position Velocity space (henceforth PPV), 
presented in LP00. One can also
study fluctuations along the {\it velocity coordinate}.  
The appropriate formulas
that relate the spectra along the velocity axis in the PPV volume to the
underlying velocity spectrum were derived in the Appendix of LP00. However,
the utility and big advantages of such a 
study  have been realized only recently. The corresponding technique was termed
the Velocity Coordinate Spectrum (henceforth VCS) 
in Lazarian (2004) and the first examples of the 
practical application of VCS are presented in Lazarian (2005, 2006).

We feel that the importance 
of this new technique calls for a more rigorous analytical study. 
Therefore one of the
 goals of the present paper is to provide more solid mathematical foundations
for the VCS technique. Even more important is to find out to what extent
the absorption affects the VCS.
 As the absorption takes place in
real space, we derive our formulas in  real space while the study
 in LP00 was done in the Fourier space.

The structure of our paper is as follows.
In \S 2 we introduce the central object of our study, namely, the 1D 
correlation
function of PPV intensities in the presence of absorption. We consider
both high resolution (pencil beam) and
finite resolution. In \S 3 we relate the correlations
of velocity and density in real space and the correlations in PPV space.
We revisit Lazarian \& Pogosyan (2004, henceforth LP04)
 and provide an improved discussion of the dominance of
velocity and density correlations to PPV correlations.
In addition, we define transformations that reveal hidden symmetries
in the 
PPV space and extend the asymptotics of PPV correlations that we obtained 
in our earlier studies. In \S 4 we consider the case of an optically thin
medium and obtain the expressions for the VCS for both narrow and wide
beams. We discuss the transition from
one regime to another as we probe different spatial scales of turbulence.
In \S 5 we derive the criteria for the VCS to be applicable to the 
observational data with self-absorption. A comparison of the VCS with other
techniques of turbulence studies, a brief discussion of the simplifications
of the model, as well as an outline of the prospects of the technique are
given in \S 6. The summary is provided in \S7. Appendixes are important parts
of our paper.
The list of our notations is given in Appendix~\ref{App:A}. 
We discuss the practical averaging of spectral line data in 
Appendix~\ref{App:B}. The fundamentals of the  PPV statistics, i.e. 
PPV correlation functions and spectra, are derived in Appendix~\ref{App:C}.  
A discussion of PPV power spectra and
the transformation that reveals the symmetries between the spatial and velocity
coordinates is provided in Appendix~D. 

\section{Correlations along Velocity Coordinate in PPV}

\subsection{The Problem: Simplified Approach}

The main object of our present study is a volume of turbulent gas or
plasma (a ``cloud''). Turbulent motions of the gas, and its
inhomogeneous distribution lead to fluctuations of intensity in the
observed Doppler shifted emission or absorption lines.
Our goal is to relate the statistical measures
that can be obtained through spectral line observations
to the underlying properties of the turbulent cascade. 
In what follows, we
concentrate on the case of the emission study, while keeping in mind that
the general formalism presented in this paper
can be easily extended to absorption studies.

We assume that the cloud extent along the line of sight $S$ is much smaller
than the distance from the volume to the observer.  This allows us 
to use the geometry of parallel lines of sight. 
Each line of sight can
then be labeled by a two-dimensional position vector ${\bf X}$ on the cloud
image, which
together with $z$ coordinate along the line of sight specifies
the three-dimensional position vector ${\bf x}=({\bf X},z)$.
Henceforth, following the convention adopted in LP00 and LP04 
we denote by the capital bold letters
the two dimensional position-position vectors reserving 
small bold letters for vectors of three dimensional
spatial position. Replacing $z$ coordinate by the observationally available
$z$-component of gas velocity gives us a vector $({\bf X},v)$ in PPV cube. 
(Where the convergence of lines of sight is essential,
cf. Chepurnov \& Lazarian, 2006, $v$ should be treated as the radial
coordinate.)

In this paper we  study the fluctuations in PPV space in
the velocity coordinate along the fixed line of sight, taking
into account the effects of self absorption of gas and the finite angular 
resolution of the telescope.  
Let us first introduce the problem in
a simplified form, disregarding the
effects of telescope resolution (cf. \S 2.2, \S 6.2.2). 
A possible statistical measure of the fluctuations of
emission intensity $I_{\bf X}(v)$  is 
the structure function 
\begin{equation}
{\cal D}({\bf X}, v_1,v_2)\equiv \left\langle \left[ I_{\bf X}(v_1) -
I_{\bf X}(v_2)\right]^2\right\rangle~~~,
\label{dv_emiss}
\end{equation}
which is the variance of the difference between intensities at two
velocities $v_1,v_2$ along the same line of sight ${\bf X}$. 
We stress that the measure (\ref{dv_emiss}) is available for the
{\it  infinite} spatial
 resolution of an instrument, while for the finite resolution
an averaging over the neighboring lines of sight with the instrumental beam
should be performed (see \S 2.2).

The intensities $I_{\bf X}(v)$ are affected by both the turbulence and
the absorption. As in LP04, to quantify these effects
consider the standard equation of radiative transfer (Spitzer 1978)
\begin{equation}
dI_{\nu}=-g_{\nu} I_{\nu} ds+j_{\nu}ds~,~~ ds=-dz ~~~,
\label{transfer1}
\end{equation}
with the absorption coefficient  
$g_{\nu}=\alpha({\bf x}) \rho({\bf x}) \phi_v({\bf x})$ and the emissivity
$j_{\nu}=\epsilon \rho({\bf x}) \phi_v({\bf x})$ are taken proportional
to the density of atoms with velocity $v$ that corresponds
to the frequency $\nu$. This density is given by
$\rho({\bf x}) \phi_v({\bf x})$,
where $\rho({\bf x)}$ is the spatial density of atoms and 
$\phi_v({\bf x})$ is the fraction of atoms that have velocity $v$.
The turbulent motions affect
the velocity distribution. Indeed,
the line-of-sight velocity $v$ of the atom 
at the position ${\bf x}$ is a sum of $z$-components of the
regular gas flow (e.g., due to galactic rotation) $v_{gal}({\bf x})$,
the turbulent velocity $u({\bf x})$ and the residual component
due to thermal motions. This residual thermal
velocity $v-v_{gal}({\bf x})-u({\bf x})$ has a Maxwellian distribution,
so
\begin{equation}
\phi_v({\bf x}) {\mathrm d} v =\frac{1}{(2\pi \beta)^{1/2}}
\exp\left[-\frac{(v-v_{gal}({\bf x})-u({\bf x}))^2}
{2 \beta }\right] {\mathrm d} v ~~~,
\label{phi}
\end{equation}
where $\beta=\kappa_B T /m_a$, $m_a$ being the mass of atoms.
If the temperature tends to zero, the function becomes a $\delta$-function,
that prescribes the velocity distribution that is determined by non-thermal
velocities. 

As shown in LP04 the solution of Eq.~(\ref{transfer1}) is 
\begin{equation}
I_{\bf X}(v)=\epsilon \int^{\rho_s}_0 dY_v {\mathrm e}^{-\alpha Y_v}=
\frac{\epsilon}{\alpha}\left[1-{\mathrm e}^{-\alpha \rho_s({\bf X},v)}\right]~~~,
\label{simplified}
\end{equation}
where
\begin{equation}
\rho_s({\bf X},v) \equiv\int_{0}^S \rho(z')\phi_v(z')dz' ~~.
\label{eq:rhos}
\end{equation}
is the density of {\it images} of the emitting atoms in PPV space, which 
henceforth we shall refer to as the PPV density. Again, following
the convention in LP00 and LP04, we use the subscript $s$ to distinguish 
the quantities in $({\bf X}, v)$ coordinates from those in  $({\bf X}, z)$
 coordinates.
As the result, Eq.~(\ref{dv_emiss}) can be written as
\begin{equation}
{\cal D}(v_1,v_2)= \frac{\epsilon^2}{\alpha^2}\left\langle 
\left[ e^{-\alpha \rho_s\left({\bf X},v_1\right)} -
e^{-\alpha \rho_s\left({\bf X},v_2\right)} \right]^2\right\rangle~~~,
\label{dv_PPV}
\end{equation}
where, for the sake of simplicity,
we omitted the label ${\bf X}$. This is what we shall do for the rest of
the paper, wherever this does not cause a confusion.

The  effects of turbulence are imprinted on ${\cal D}(v_1,v_2)$
through $\rho_s({\bf X},v)$.
The PPV density  $\rho_s({\bf X},v)$ 
depends on the real density of gas $\rho({\bf X}, z)$ and 
on velocity of gas particle $v({\bf X}, z)$.

\subsection{Effects of Finite Angular Resolution}
Realistic observations have a finite angular resolution.
The emission intensity measured by a telescope is 
$\int d{\bf X_1} B({\bf X_1}) I_{\bf X_1} (v_1)$, 
where $B({\bf X_1}) $ is the beam of the instrument, assumed to be centered 
on the line of sight at ${\bf X_1}=0$. The explicit introduction of the
beam was avoided in LP00 and LP04 as those papers dealt with slices
of data for which spatial resolution was essential. For the case of
studies of fluctuations along the velocity coordinate, meaningful
results may be obtained for spatially unresolved eddies as well. 

Since integration commutes with taking ensemble average,
the structure function of the beam-smeared signal
is given by
\begin{equation}
{\cal D}(v_1,v_2)\equiv 
\int d {\bf X_1} B({\bf X_1}) \int d {\bf X_2} B({\bf X_2})
\left\langle 
\left[ I_{\bf X_1}(v_1) - I_{\bf X_1}(v_2)\right]
\left[ I_{\bf X_2}(v_1) - I_{\bf X_2}(v_2)\right]
\right\rangle~~~,
\label{dv_emiss_beam}
\end{equation}
that is
a generalization of Eq.~(\ref{dv_emiss}).

Using Eq.~(\ref{simplified}) and a short hand notation 
$\rho_{ij}=\rho_s({\bf X}_i,v_j)$, we
can express ${\cal D}(v_1,v_2)$ via PPV density in the following form
\begin{eqnarray}
{\cal D}(v_1,v_2)&=&  \frac{\epsilon^2}{\alpha^2}
\int d {\bf X_1} B({\bf X_1}) \int d {\bf X_2} B({\bf X_2}) \times
\nonumber \\ &&
\langle e^{-\alpha (\rho_{11}+\rho_{21})}
\left[  1 
       -e^{-\alpha (\rho_{12}-\rho_{11})}
       -e^{-\alpha (\rho_{22}-\rho_{21})}
       +e^{-\alpha (\rho_{12}-\rho_{11}+\rho_{22}-\rho_{21})}
       \right]
\rangle~~.
\label{dv_PPVgen}
\end{eqnarray}
The idea behind writing the structure function in this form 
is to notice that the term in square brackets
depends on the difference between PPV densities
at different velocities, but along the same line of sight, 
$\rho_{21}-\rho_{22}$ and $\rho_{11}-\rho_{12}$, while the
prefactor, $\rho_{11}+\rho_{21}$, is evaluated at the same velocity, 
although between different lines of sight.
This makes Eq.~(\ref{dv_PPVgen}), still fully general,
particularly convenient for studying scaling behavior at small
velocity separations.

To proceed further we need to consider
some limiting cases and approximations. 
At sufficiently small $v=v_1-v_2$, one can expand the terms in the brackets
into the power series
\begin{eqnarray}
{\cal D}(v)&\sim&  \epsilon^2
\int d {\bf X_1} B({\bf X_1}) \int d {\bf X_2} B({\bf X_2}) \times
\nonumber \\ &&
\langle e^{-\alpha (\rho_{11}+\rho_{21})}
\left[ (\rho_{12}-\rho_{11})(\rho_{22}-\rho_{21}) 
       + O\left(\alpha \Delta \rho^3\right)
       + O\left(\alpha^2 \Delta \rho^4\right)
       \right]
\rangle~~,
\label{dv_PPVexpand}
\end{eqnarray}
where we assumed homogeneity of correlation functions in the velocity
direction. As we discuss in Appendix~\ref{App:B} this is not a necessary condition for
the theory formulation, but we use it to simplify both our discussion and
notations.
 
In LP04
\footnote{Our consideration is similar to the one we advanced in LP04
for the two dimensional structure function of intensity. There, however,
the role of velocity $v$ and angular $R$ directions were reversed.}
we argued, that factoring out the averaging over the ``absorption window''
$e^{-\alpha (\rho_{11}+\rho_{12})}$ 
\begin{eqnarray}
{\cal D}(v)&\sim&  \epsilon^2
\int d {\bf X_1} B({\bf X_1}) \int d {\bf X_2} B({\bf X_2}) \times
\nonumber \\ &&
\langle e^{-\alpha (\rho_{11}+\rho_{21})} \rangle
\left[ \langle  (\rho_{12}-\rho_{11})(\rho_{22}-\rho_{21}) \rangle
       + \langle O\left(\alpha \Delta \rho^3\right) \rangle
       + \langle O\left(\alpha^2 \Delta \rho^4\right) \rangle
       \right] ~~.
\label{dv_PPVexpfac}
\end{eqnarray}
provides a good approximation for studying the onset
of the absorption effects. 

In the case of an isotropic beam and homogeneous statistics the last 
expression becomes
\begin{equation}
{\cal D}(v)\sim  \epsilon^2
\int R dR \;B^2(R) \;  W_{abs}(R)
\left[ d_s(R,v)-d_s(R,0) 
       + \langle O\left(\alpha \Delta \rho^3\right) \rangle
       + \langle O\left(\alpha^2 \Delta \rho^4\right) \rangle
       \right] ~~,
\label{dv_PPVexpfin}
\end{equation}
where $R=|{\bf X_1}-{\bf X_2}|$, 
the PPV structure function is
\begin{equation} 
d_s(R,v) = \langle
        \left[ \rho_s({\bf X_2},v_2)-\rho_s({\bf X_1},v_1) \right]^2 \rangle
~~~,
\label{dsrv}
\end{equation}
the absorption window $W_{abs}$ is
\begin{equation}
W_{abs}(R) = \langle 
  e^{-\alpha \left[ \rho_s({\bf X_1},v_1)+\rho_s({\bf X_2},v_1)\right] }
  \rangle ~~~,
\label{Wabs}
\end{equation}
and
\begin{equation}
B^2(R) = 2 \pi \int d{\bf X}_+ B({\bf X}_+ + {\bf R}/2)
B\left({\bf X}_+ - {\bf R}/2\right), ~~~{\bf X_+}=({\bf X}_1 + {\bf X}_2)/2~~.
\end{equation}
Eq.~(\ref{dv_PPVexpfin})
provides the foundations of the formalism for finite resolution studies.

\subsection{High Resolution Limit}

The formalism is significantly simplified in the limit of infinitely high 
angular resolution, $B({\bf X})=\delta({\bf X})$. The criterion for neglecting
the beam spread depends
on the properties of the turbulence. 

For infinitely high (henceforth, {\it high})
 resolution Eq.~(\ref{dv_PPVgen}) becomes 
\begin{eqnarray}
{\cal D}(v) &=& \frac{\epsilon^2}{\alpha^2}
\langle e^{-2 \alpha \rho_{1}} 
\left[  1 - e^{-\alpha (\rho_{2}-\rho_{1})} \right] ^2
\rangle \nonumber \\
&\approx&  \frac{\epsilon^2}{\alpha^2} 
\langle e^{-2 \alpha \rho_{1}} \rangle
\langle
\left[  1 - e^{-\alpha (\rho_{2}-\rho_{1})} \right] ^2
\rangle~~.
\label{dv_PPVideal}
\end{eqnarray}
where we have omitted a now irrelevant first spatial
index in $\rho$, $\rho_i\equiv\rho(0,v_i)$.
At the velocity scales both large enough for the beam width
to be neglected (which allows to set $R$ to zero),
and small enough for the series expansion to be accurate
\begin{equation}
{\cal D}(v)\sim  \epsilon^2 W_{abs}(0)
\left[ d_s(0,v)
       + \alpha \langle (\rho_1-\rho_2)^3 \rangle
       + \frac{\alpha^2}{12} \langle (\rho_1-\rho_2)^4 \rangle
       + O(\alpha^4)
\right] ~~.
\label{dv_PPVidexp}
\end{equation}

The Eqs~(\ref{dv_PPVexpfin}), (\ref{dv_PPVidexp}) provide the basis
for the subsequent discussion of the
short scale limit  of statistical descriptors and the conditions under which
the absorption start to play a role.
They establish
the link between observable fluctuations of intensity 
 along the
line of the velocity coordinate, 
which are characterized here by 
${\cal D}(v)$, and statistical descriptors of PPV density
$\rho_s$, namely, $d_s(R,v)$ (see Eq.~(\ref{dsrv})).

\section{Fluctuations in Real Space and PPV Space}

This section presents the ground work in relating the correlation and structure
functions in the PPV space and the relevant statistical descriptors of the
velocity and density in the emitting turbulent volume.

\subsection{Statistical measures of velocity, density and PPV density}

\subsubsection{Correlations and spectra in real space}

It is well known that in the presence of magnetic field turbulence 
becomes axisymmetric
in the reference frame related to the {\it local} direction
of magnetic field (see discussion in Cho, Lazarian \& Vishniac  2002
and references therein). However, as we have discussed earlier 
(see LP00, LP04),
to a large extent it is possible to use isotropic statistics of
density and velocity when dealing with observations. Briefly,
this is related to the fact that the observations are performed
in the {\it global} system of reference related to the mean magnetic
field. In this system of reference the anisotropy is rather mild and
spectra in directions parallel and perpendicular to the magnetic field
have the same scaling. These considerations were successfully tested
using synthetic observations with the data cubes obtained through
direct 3D MHD simulations (Esquivel et al. 2003).    

The fact above permits us to use the standard isotropic statistical measures
like structure function, correlation function and spectra
(see Monin \& Yaglom 1976) to describe both the velocity and density 
correlations in the turbulent cloud under study. In what follows, we use
the statistically isotropic (see Lazarian 1995 for a more general case)
correlation function of the $xyz$-space density field $\rho({\bf x})$ 
\be
\xi(r)=\xi({\bf r}) = \langle \rho ({\bf x}) \rho ({\bf x}+{\bf r}) \rangle~~
~,
\label{xifirst}
\ee
as well as
 the correlation function of the density fluctuations
$\delta\rho=\rho-\langle\rho\rangle$
\be
\tilde\xi(r)= \langle \delta\rho ({\bf x})
\delta\rho ({\bf x}+{\bf r})\rangle = \xi(r)-\langle\rho\rangle^2
~.
\label{tildexi}
\ee
At zero lag, $\xi(0)=\langle \rho^2 \rangle$ is the second moment of the
field.  At large separations the correlation function
provides the square of the mean value, $\xi(\infty) \to {\bar \rho}^2$,
while $\tilde\xi(\infty) \to 0$.

In addition, we use the density structure function
\begin{equation}
d(r)= \langle(\rho({\bf x}+{\bf r})-\rho({\bf x}))^2 \rangle~~~,
\label{2}
\end{equation}
and the density power spectrum
\begin{equation} 
P({\bf k})=\int d {\bf r} e^{i{\bf k}{\bf r}} \xi({\bf r}) ~~or~~
P({\bf k})=-\onehalf \int d {\bf r} e^{i{\bf k}{\bf r}} \d({\bf r})~~~.
\label{spectrum:gen}
\end{equation}

The utility of using both structure and correlation functions can be illustrated
with density fluctuations. 
An essential difference between the two is that, while the value
of the structure function at some scale $r$, $d(r)$,
is determined by the integrated power of fluctuations
over smaller scales $r^\prime \le r$,
the value of the correlation
function $\xi(r)$ reflects the integral of the power
over scales $r^\prime \ge r$.
Therefore, the correlation functions are more 
appropriate to use for spectra where most power is at increasingly 
small scales (Monin \& Yaglom 1975), in particular
for power-law spectra $P(k)\sim k^{n}$, for $n>-3$.
Following LP00 and LP04,
we call such spectra ``shallow''. In contrast, the
``steep'' power spectra, for which $n<-3$ and most of the
power is on the large scales, are more robustly described by
the structure function.

For exact power law spectra with steep index, the correlation function is not
formally well defined due to divergent contribution
from large scales
\footnote{For very steep spectra, i.e. for $n<-5$, even
structure functions fail to keep in check the contribution
from large scale gradients and to properly reflect the
turbulence statistics (Monin \& Yaglom, see also
examples in Cho \& Lazarian 2004, 2005). We obtain such
steep power spectra for fluctuations of intensity along the
velocity coordinate and the problems that this entails are addressed below.},
while for shallow index the structure function diverges due to small scale 
power.
However, as we discussed in LP04,
if the appropriate cut-offs at small and/or large scales are introduced,
both descriptors can be used simultaneously and relate simply to each other
\begin{equation}
d(r) = 2 \left[\xi(0) - \xi(r)\right]~,~~ \tilde \xi(r) = 
\onehalf \left[ d(\infty)-d(r)\right] ~~.
\label{eq:d_xi}
\end{equation}
In this case $d(\infty)=2\left(\xi(0)-\bar\rho^2\right)=2 \tilde\xi(0)$.

For a shallow power law spectrum the correlation function is a decaying power
law $\tilde \xi (r) \propto r^{-\gamma}$ with $\gamma > 0$, while for a steep 
power-law spectrum the structure function $d(r) \propto r^{-\gamma}$,
is a rising power-law with $\gamma < 0$ (see Monin \& Yaglom 1975, and numerical
examples in Esquivel \& Lazarian 2005).  
The relation between the spectral index $n$ of $P(k)$ and the index $\gamma$ 
of structure (for a steep spectrum) or correlation (for a shallow spectrum)
functions is straightforward:
\begin{equation}
{\rm spectral~ index}= \gamma-{\rm dimensionality~of~space}~.
\end{equation}
In this notation Kolmogorov turbulence in 3D has $\gamma=-2/3$ 
and the spectral index of $P(k)$,  $n=-11/3$. The usually quoted
Kolmogorov $-5/3$ index corresponds to the direction averaged spectrum, 
namely to  $4\pi k^2 P(k)$.

We also use a structure function to characterize the turbulent velocity.
Since only the $z$-component of the velocity field is available, we just need 
\begin{equation}
D_z({\bf r})=
\langle(u_z({\bf x}+{\bf r})-u_z({\bf x}))^2 \rangle ~~. 
\label{Dz}
\end{equation}
Unlike the density for which both shallow and steep cases
can be realized depending on the ratio of fluid viscosity
and resistivity (see Lazarian, Vishniac \& Cho 2004), and Mach number
(see Beresnyak, Lazarian \& Cho 2005), the velocity field scaling
is always steep (see Cho, Lazarian \& Vishniac 2003).
The latter we write as $D_z(r) \propto r^m, ~ m>0$.
\footnote{Note the difference in the signs of $\gamma$ that we use
for density with steep spectra and $m$ used for velocity structure functions.
This difference is somewhat unfortunate, but in case of density where both
steep and shallow spectra are possible, it allows us to
write $\sim r^{-\gamma}$ scaling universally for both the correlation and 
structure functions.}

\subsubsection{Correlations in PPV}

In what follows we assume 2D statistical homogeneity and isotropy
of $\rho_s({\bf X},v)$
in the ${\bf X}$-direction over the image of a cloud.
Homogeneity and isotropy in 
${\bf X}$ causes the mean density to depend only on velocity, while
the correlation functions depend only on the magnitude of separation between
two sky directions
$R=|{\bf R}|=|{\bf X}_1-{\bf X_2}|$.
Then, in PPV space the mean density is
\begin{equation}
\bar\rho_s(v_1)=\langle \rho_s({\bf X}_1,v_1) \rangle~~~, 
\end{equation}
and the correlation functions of the density and, closely related, density
fluctuations $
\delta\rho_s({\bf X}_1,v_1) = \rho_s({\bf X}_1,v_1) - \bar \rho_s(v_1) $
are
\begin{eqnarray}
\label{eq:xis}
\xi_s(R,v_1,v_2) &\equiv& \langle \rho_s({\bf X}_1,v_1)
\rho_s({\bf X}_2,v_2) \rangle~~, \\
\tilde \xi_s(R,v_1,v_2) &\equiv& \langle \delta\rho_s({\bf X}_1,v_1)
\delta\rho_s({\bf X}_2,v_2) \rangle = 
\xi(R,v_1,v_2) - \bar \rho_s(v_1) \bar\rho_s(v_2) ~.
\end{eqnarray}
For PPV statistics the structure functions are
\begin{eqnarray}
d_s(R,v_1,v_2)&=&\langle \left[ \rho_s({\bf X}_1,v_1)-\rho_s({\bf X}_2,v_2)\right]^2\rangle~~, \\
\tilde d_s(R,v_1,v_2)&=&\langle \left[\delta\rho_s({\bf X}_1,v_1) -
\delta\rho_s({\bf X}_2,v_2)\right]^2 \rangle 
= d_s(R,v_1,v_2) - \left[ \bar\rho_s(v_1) - \bar\rho_s(v_2) \right]^2~.
\end{eqnarray}
Here we have maintained the notation which highlights the
symmetries of the correlation and structure functions. The conditions under
which the homogeneity along a velocity direction is fulfilled are formulated in
Appendix~\ref{App:B}. For the rest of the paper we assume that they are
 satisfied and we assume that the PPV correlation and structure functions 
depend only on the velocity difference $v$. 
In Section~\ref{sec:symmetry} we discuss some asymptotic symmetries between
${\bf X}$ and $v$ directions.

\subsubsection{Relation between PPV and real space correlations}

The PPV correlations can be expressed via real space velocity
and density correlations using the set of assumptions that was employed 
in LP00 and LP04 (see also Appendix~\ref{App:C}). 
If the gas is confined in an isolated cloud of size $S$ and the
galactic shear over this scale is neglected, the zero-temperature correlation
function is (see Appendix~\ref{App:C})
\begin{equation}
\xi_s(R,v) \propto
\int_{-S}^S {\mathrm d}z \left(1-\frac{|z|}{S}\right)
\; \frac{\xi( r)}{D_z^{1/2}({\bf r})}
\exp\left[-\frac{v^2}{2 D_z({\bf r})}\right],
\label{ksicloud}
\end{equation}
where, assuming for the sake of simplicity
that the velocity field is solenoidal,
the structure function of the 
$z$-components of velocity is (see LP04) 
\begin{equation}
D_z({\bf r})=Cr^{m}\left[1+\frac{m}{2}\left(1-z^2/r^2 \right)\right]~~,
\end{equation}
where $C$ is a normalization constant.
We keep the vector notation, as, unlike the correlation function of
the scalar density field, the structure function  of even isotropic
vector field depends not just on $r$, but also on the angle with the z-axis
(via $z/r$)  (see Monin \& Yaglom
1975).  In terms of $R$ and $z$, a useful
explicit expression is
\begin{equation}
D_z({\bf r})=C \left(R^2+z^2\right)^{m/2-1}
\left( [1+m/2] R^2 + z^2 \right) ~~~.
\end{equation}
If the turbulent energy cascades
from a scale larger or equal to
the cloud size $S$,  $D_z({\bf r})$ is growing 
up to the scale of the cloud $S$ at which the
velocity structure function saturates at the value $D_z(S) 
= CS^m$.    
Note, that a
 somewhat complicated prefactor entering Eq.~(\ref{AppB:2point_cloud})
is omitted in Eq.~(\ref{ksicloud}),
which is justified as we are interested in the functional
dependence of $\xi_s$ rather than its amplitude.
This expression is obtained by substituting $\rho_s$ given
by Eq.~(\ref{eq:rhos}) into the definition of the correlation function
(\ref{eq:xis}) and performing averaging over the Gaussian distribution
of the turbulent velocity field $u({\bf x})$,
contained in $\phi_v({\bf x})$ (see Appendix~\ref{App:C}).

For the studies of turbulence in the presence of  a regular flow,
e.g. the galactic shear velocity $v_{gal}$, one should replace
$v$ in the exponent in Eq.~(\ref{ksicloud}) by $v-v_{gal}$ (see LP00 and LP04).
We found that neglecting
the regular flow provides an adequate approximation for studies
in a wide range of circumstances.
Besides isolated clouds, this includes observations in directions
of high galactic latitude, but also the case of HI in the Galactic plane,
if we focus on small scale phenomena. Indeed, the velocity gradient arising 
from the Galactic rotation is $ \sim 0.7~\mathrm{km/s/50pc}$
and decreases linearly to shorter scales.
At the same time turbulent relative motions of HI are
expected to be $20~{\mathrm km/s}$ at $50 pc$ and scale as $r^{1/3}$.
Thus, even if turbulent
scaling saturates at $S \sim 50 pc$, for considering linear scales
$r < S$, or, correspondingly, in velocity direction 
$v < D_z^{1/2}(S) \sim 20~{\mathrm km/s}$, it is justifiable to
neglect coherent shearing motions arising from galactic rotation. 
If turbulence is extended to ever larger
scales the region of validity of the approximation extends until
turbulent velocities equate with the shearing motion. This scale,
denoted $\lambda$ in LP00, is equivalent to the
``cloud'' size\footnote{
LP00 also contains asymptotics for the regime when shear-like regular motions
exceed the turbulent velocities, which can provide a starting point
for future detailed consideration of this case.} $S$.

The effect of thermal motions is described by 
the convolution of correlation functions
evaluated at a vanishing temperature with a thermal window (Appendix~\ref{App:C})
\begin{equation}
\xi_s(R,v) \propto
\int_{-\infty}^\infty
\frac{dv^\prime}{(4 \pi \beta)^{1/2}} 
\exp\left[-\frac{(v-v^\prime)^2}{4 \beta}\right]
\int_{-S}^S {\mathrm d}z \left(1-\frac{|z|}{S}\right)
\; \frac{\xi( r)}{D_z({\bf r})^{1/2}}
\exp\left[-\frac{{v^\prime}^2}{2 D_z({\bf r})}\right] ~~~.
\label{ksicloud_temp}
\end{equation}

The structure function is computed from the formal expression 
\begin{equation}
d_s(R,v) = 2 \left[\xi_s(0,0) - \xi_s(R,v) \right]
\label{dcloud}
\end{equation}
where 
\begin{equation}
\xi_s(0,0)  \propto 
\int_{-S}^S {\mathrm d}z \left(1-\frac{|z|}{S}\right)
\; \frac{\xi( z)}{D_z({z})^{1/2}} ~~~,
\end{equation}
Frequently, the two terms on the right hand side of Eq.~(\ref{dcloud})
are individually divergent, but their combination is not, provided
that the subtraction is done before the integration over $z$.

\subsection{Fluctuations of Density: Correlation Radius $r_0$}

\subsubsection{Shallow density spectrum}
For shallow density spectra, we use power-law correlation functions
of {\it overdensity}:
\begin{equation}
\xi(r)= \langle \rho \rangle^2 
\left(1 + \left[ {r_0 \over r} \right]^\gamma\right), ~~~~~ \gamma > 0~~~
\label{Appeq:xi}
\end{equation}
where $r_0$ has the physical meaning of the scale at which fluctuations
are of the order of the mean density. 
This ansatz describes properly the situation where the deviations from the
mean become uncorrelated at large distances. 
We have argued in LP04 that the amplitude
of density perturbations at the scale of a cloud should not exceed the mean
density, which for the shallow density spectrum translates to the requirement
that $r_0<S$.

\subsubsection{Steep density spectrum}
To describe correlations in the random density field
with steep (i.e. $n<-3$) power spectrum we start with 
the structure function given by Eq.~(\ref{2}).
Real world structure functions do not grow infinitely and therefore
there is a cut-off at some large scale $r_c$ above which the structure
function saturates at some limiting value. Being
interested in $r \ll r_c$ we shall not address here the 
issue of the precise form of the saturation and will proceed with
a simple ansatz
\begin{equation}
d(r)=d(\infty){r^{-\gamma} \over r^{-\gamma}+r_c^{-\gamma}}~~,~~ \gamma < 0 ~~~.
\label{structure}
\end{equation}
To find the characteristic correlation length in this case, we note that
the correlation function (now well defined because of the 
cutoff at large scales  so that Eq.~(\ref{eq:d_xi}) can be used) is
\begin{equation}
\xi(r)=
\frac{1}{2} d(\infty)[1-d(r)/d(\infty)] + \langle \rho \rangle^2
=\frac{d(\infty)}{2}{r_c^{-\gamma} \over r^{-\gamma}+r_c^{-\gamma}}
+ \langle \rho \rangle^2 ~~~.
\label{cor-structure}
\end{equation}
At sufficiently small $r\ll r_c$ Eq.~(\ref{cor-structure}) gives
\begin{equation}
\xi(r)\approx \left(\frac{1}{2} d(\infty)+\langle \rho \rangle^2 \right)
\left(1- 
\frac{d(\infty)}{d(\infty)+2 \langle \rho \rangle^2}[r/r_c]^{-\gamma} \right)~~.
\label{xinew}
\end{equation}
Introducing the correlation scale 
\begin{equation}
r_0=
r_c \left[1+2 \langle \rho \rangle^2 /d(\infty)\right]^{-\frac{1}{\gamma}}
\label{r0}
\end{equation}
we recast Eq.~(\ref{xinew}) in a form similar to Eq.~(\ref{Appeq:xi})
\begin{equation}
\xi(r)\approx \left(\frac{1}{2} d(\infty)+\langle \rho \rangle^2 \right)
\left(1- \left[\frac{r_0}{r}\right]^{\gamma} \right)~~,~~~ \gamma<0
\label{xinewr0}
\end{equation}
which allows the uniform treatment of both the steep and shallow cases.

\subsubsection{Physical meaning of $r_0$}
For a fixed $\gamma$ 
the density correlation scale $r_0$ determines the amplitude of
the density fluctuations at a scale $r$
relative to the term that plays the role of the mean uniform density.
While for the shallow spectra the uniform density in the volume
is just the ensemble average
density $\langle \rho \rangle$ (see Eq.~(\ref{Appeq:xi})), for the steep spectra the role
of the uniform density factor is played by 
$\sqrt{\langle \rho \rangle^2+d(\infty)/2}$ (see Eq.~(\ref{xinewr0})). The meaning is clear: 
since the high amplitude 
density perturbations with the steep spectra are concentrated at
the largest scales $\sim r_c$,
in a relatively small volume all long-wave modes
give mostly uniform offsets of the density, forming a background
on which we study small scale ripples. In other words, they serve as local
mean densities. The typical 
magnitude of these modes is described by the dispersion,
$\frac{1}{2} d(\infty)$.
Besides that, there is a contribution arising from the overall global 
mean density $\langle \rho \rangle^2$.
As the result, $r_0 \ge r_c$ always (see Eq.~(\ref{r0})).

\subsection{Velocity \& density: Revisiting LP04}

The term 
$(1\pm (r_0/r)^\gamma)$ in the expressions of the density correlation
functions given by Eqs.~(\ref{xinewr0}) and (\ref{Appeq:xi})
results in the separation of the of the PPV correlation function 
into two parts (LP04)
\begin{eqnarray}
\tilde \xi_s(R,v) &=& \tilde \xi_v(R,v)
        + \tilde\xi_\rho(R,v), \nonumber \\
\tilde d_s (R,v) &=& \tilde d_v(R,v)+ \tilde d_\rho (R,v). 
\label{eq:rho_v_split}
\end{eqnarray}
with the $v$-term describing pure velocity effects, while the $\rho$-term
arises from the actual real space density inhomogeneities that are
modified
by velocity mapping. To simplify the notation we have dropped the index
$s$ from the right-hand-side quantities, since this split is only meaningful
in PPV space. This also corresponds to the split in the PPV spectra discussed
in LP00.

In LP04 we defined the amplitude of density perturbations for steep spectrum
in a different way, which resulted in some confusion, which, fortunately, did
not affect our final results there. Here we present a correct discussion of the
contribution of the density and velocity fluctuations to the amplitude of
fluctuations in PPV.

In PPV space
 correlation and structure
functions are split into two contributions
according to Eq.~(\ref{eq:rho_v_split}),
it is possible to see that the term that depends only on velocity fluctuations
originates from the effective mean density part in correlation functions
(\ref{Appeq:xi}) and (\ref{xinewr0}), while the $\rho$-term
arises from the density fluctuation part that is modified by velocity 
fluctuations.
The scale $r_0$ is, thus,
critical for establishing their relative magnitudes.
Namely, for shallow density spectra (i.e. $\gamma > 0$)
the pure velocity effect dominates the density fluctuations at large $r > r_0$,
while for the steep density spectra (i.e. $\gamma < 0$)  the pure velocity term
dominates at small scales, $r < r_0$. Since as we argued in \S 3.2.3 
in the latter case $r_0 > r_c$, the  density fluctuations  are never dominant
for $\gamma<0$. This discussion invalidates
the results in the second column in Table~2 in LP04, where conditions for 
the dominance of density fluctuations are formulated for $\gamma<0$. 
The density term can dominate only if $\gamma > 0$, the regime which is
summarized below in the revised Table~\ref{table:density}.
\begin{table}[ht]
\centering
\begin{tabular}{llc}
Condition & $ \gamma > 0 $ & Eqns \\
$ m \ge \mathrm{max}\left[\frac{2}{3},\frac{2}{3}(1-\gamma)\right] $
 & $ v^2 <  D_z(S) (r_0/S)^m $ &
(\ref{AppCeq:d_rhoVf1})-(\ref{eq:d_vVfg23})\\
$ \frac{2}{3}(1-\gamma) < m < \frac{2}{3} $
& $ v^2 < D_z(S) (r_0/S)^{\frac{2/3 \gamma m}{m-2/3(1-\gamma)}} $ &
(\ref{AppCeq:d_rhoVf1})-(\ref{eq:d_vVfl23})\\
$ m \le \mathrm{min}\left[\frac{2}{3},\frac{2}{3}(1-\gamma)\right] $
& $ r_0/S > 1 $ &
(\ref{AppCeq:d_rhoVf2})-(\ref{eq:d_vVfl23})
\end{tabular}
\caption{Range of the scales where the impact of
density inhomogeneities to the PPV statistics exceeds
the velocity contribution.
}
\label{table:density}
\end{table}

In terms of the practical interpretation of observations
 our present finding allows a
more reliable interpretation of the power spectra in terms of underlying 
velocity fluctuations. Indeed, one should not worry about density
contamination of the results of both VCA and VCS
if the underlying density spectrum is steep. 

\subsection{Asymptotics of line-of-sight PPV correlations}
\label{sec:losdv}
The line-of-sight correlations are of particular importance for our
present study. The expression for 
$\tilde d_{\rho}(0,v)$ is related to the PPV correlation function as
\begin{equation}
\tilde d_{\rho}(0,v)=2\left[\xi_\rho(0,0)-\xi_\rho(0,v)\right]~~~.
\label{relation}
\end{equation}
A similar expression is valid for $\tilde d_{v}(0,v)$. For the power-law
small-scale statistics that we deal with in this paper $\tilde d_{v}(0,v)$
can be obtained from $\tilde d_{\rho}(0,v)$ by setting $\gamma=0$.
Therefore, without losing generality, we shall consider
$\tilde d_{\rho}(0,v)$ only, which is given by the integral (see Eqs.~(\ref{ksicloud}) and (\ref{relation}))
\begin{eqnarray}
\tilde d_{\rho}(0,v) &\sim&
\left(\frac{r_0}{S}\right)^\gamma 
\int_{-1}^1 {\mathrm d}\hat z \;
\frac{1}{|\hat z|^{\gamma+m/2}} \left[ 1 -
\exp\left(-\frac{{\hat v}^2}{2 |\hat z|^m}\right) \right] \nonumber \\
&\propto& \frac{{\bar \rho}^2 S^2}{ D_z(S)} \frac{1}{m}
\left(\frac{r_0}{S}\right)^\gamma 
\left[ \frac{1}{p} - \left(\frac{{\hat v}^2}{2}\right)^p
\Gamma\left(-p,\frac{{\hat v}^2}{2}\right) \right]
\label{eq:d_vV}
\end{eqnarray}
where $p=(1-\gamma)/m-1/2 > 0$ and,
to shorten intermediate formulas, the dimensionless quantities
$\hat v=v/D_z^{1/2}(S)$, $\hat z=z/S$, $\hat r = r/S$ are introduced.
Here $\Gamma$ is the incomplete gamma-function, but
when $\Gamma$ is used with one argument, the ordinary gamma function is
implied.

If we perform the series expansion of the incomplete gamma-function for
small argument $\hat v$, the square brackets in Eq.~(\ref{eq:d_vV})
provide
\begin{equation}
p^{-1} - 2^{-p} {\hat v}^{2p} \Gamma\left(-p,{\hat v}^2/2\right) 
= - 2^{-p} {\hat v}^{2p}\Gamma[-p] -
\left( \frac{1}{2(1-p)} {\hat v}^2 + O({\hat v}^4)\right)~~.
\label{eq:Gamma}
\end{equation}
The informative part of the structure function is
in the ${\hat v}^{2p} \Gamma[-p]$ term, while terms in the series part 
of Eq.~(\ref{eq:Gamma}) individually have
no information about the underlying scaling and thus present a 
problem for a straightforward use of structure functions. 
For instance, the very first $v^2$ term in the series
is dominant
whenever $p > 1$, saturating the structure function at $v^2$ behavior.
\footnote{The series expansion for the incomplete gamma-function is irregular
at integer $p$. A rigorous treatment of these cases, most notable of which is
the case of Kolmogorov velocity, $m=2/3,\gamma=0,p=-1$, gives rise to
additional logarithmic factors.}
More precisely, the leading behavior of the structure function given
by Eq.~(\ref{eq:d_vV}) is
\begin{eqnarray}
{\tilde d}_{\rho}(0,v) &\propto& \frac{{\bar \rho}^2 S^2}{ D_z(S)} 
\left(\frac{r_0}{S}\right)^\gamma \; A_{0v}(\gamma,m)
\left(\frac{v^2}{D_z(S)}\right)^{\frac{1-\gamma}{m}-\frac{1}{2}}~,~ 
m > \frac{2}{3} (1-\gamma)~,
\label{AppCeq:d_rhoVf1} \\
{\tilde d}_{\rho}(0,v) &\propto& \frac{{\bar \rho}^2 S^2}{ D_z(S)}
\left(\frac{r_0}{S}\right)^\gamma \;
A_{0v}(\gamma,m) \; \frac{v^2}{D_z(S)}
~~~~~~~~~~~~~~,~ m < \frac{2}{3} (1-\gamma)~,
\label{AppCeq:d_rhoVf2}
\end{eqnarray}
with numerical factors $A_{0v}(\gamma,m)$ given in Appendix~\ref{App:A}.
In particular, setting $\gamma=0$ for the $v$-term contribution,
\begin{eqnarray}
{\tilde d}_v(0,v) &\propto& \frac{{\bar \rho}^2 S^2}{ D_z(S)} \;
A_{0v}(0,m) 
\left(\frac{v^2}{D_z(S)}\right)^{\frac{1}{m}-\frac{1}{2}}~,
~~~~~~~~~~~~~~ m > \frac{2}{3}~,
\label{eq:d_vVfg23} \\
{\tilde d}_v(0,v)&\propto& \frac{{\bar \rho}^2 S^2}{ D_z(S)} \;
A_{0v}(0,m) \; \frac{v^2}{D_z(S)} ~,~~~~~~~~~~~~~~~~~~~~~~~
~m < \frac{2}{3}~,
\label{eq:d_vVfl23}
\end{eqnarray}
Notably, for the pure velocity effect,
the value $m=2/3$, which corresponds to Kolmogorov
velocity scaling, presents a boundary case.

Where does the saturation come from? The $v^{2p}$ contribution arises from
small spatial separations $z \sim 0 $ in the integral in Eq.~(\ref{eq:d_vV})
reflecting small scale turbulent behavior, while $v^2$  terms come from
the longest separations $z \sim S$.  Hence, we see that when $p > 1$ 
the longwave modes of the size of the cloud dominate
the velocity structure function even at small velocity separations $v$ and mask
the information about the small spatial scales.
We therefore have to conclude that the two-point velocity coordinate 
structure function is not a useful quantity to measure {\it directly},
unless $m > 2/3(1-\gamma)$, which, in particular, excludes
the case when Kolmogorov turbulent velocity dominates the fluctuations
of intensity.

However, as we show in \S 4.1 , 
the terms that saturate the structure function
only provide the contribution to the power spectra 
that is localized at long wavelengths.
Therefore in the Fourier domain one can evaluate the short-wave spectrum
of the PPV fluctuations along the velocity coordinate for a more
general set of $m$ and $\gamma$ values.  

\subsection{Hidden symmetry between line-of-sight and angular PPV correlations}
\label{sec:symmetry}
The Eq.~(\ref{dv_PPVexpfin})
requires the knowledge of $d_s(R,v)$ in both positional,
$R$, and velocity, $v$, directions.
In LP04 we obtained
\begin{equation}
\tilde d_\rho(R,0)
\propto -\frac{{\bar \rho}^2 S^2 (r_0/S)^\gamma}{ D_z(S)} A_{R0}(\gamma,m)
\left(\frac{R}{S}\right)^{1-\gamma-m/2}  
\label{AppC:eq_drhoR}
\end{equation}
$A_{R0}(\gamma,m)$ is defined as an integral when the angular dependence of
$D_z({\bf r})$ is taken into account (see Appendix~\ref{App:A}). It reduces
to the combination of Gamma functions of LP04 if this dependence is ignored.

Here we want to present a remarkable symmetry between the line-of-sight
statistics and the statistics in PP direction. Indeed, the comparison
of Eq.~(\ref{AppCeq:d_rhoVf2}) and Eq.~(\ref{AppC:eq_drhoR}) shows that if one
identifies with velocity $v$ a linear scale $\ae$ according\footnote{We 
intentionally use a the unusual symbol $\ae$ to stress the peculiar non-linear
nature of the coordinate transformation that reveals the hidden symmetries in
the PPV. Although $\ae$ has dimensionality of a distance, one should not 
confuse it with a real distance to an emitter.} to
\begin{equation}
v^2 = F(\gamma,m) D_z(S) \left(\ae/S\right)^m ~~~,
\label{eq:vzmap}
\end{equation}
where 
$F(\gamma,m)=\left[A_{R0}(\gamma,m)/A_{0v}(\gamma,m)\right]^{m/(1-\gamma-m/2)}$,
then the line-of-sight structure function $d_s(0,\ae)$ acquires the
asymptotic form, identical to $d_s(R,0)$, 
\begin{equation}
\tilde d_s(0,t) = \tilde d_s(t,0)
\propto \left( \frac{t}{S} \right)^{1-\gamma-m/2} ~~~.
\end{equation}

\begin{figure}[ht]
\begin{center}
 \includegraphics[width=0.5\textwidth]{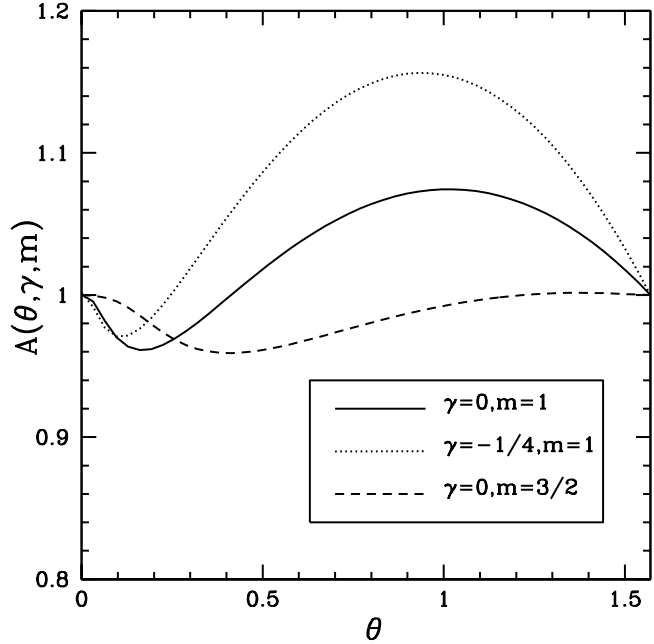}\\
\end{center}
\caption{Function $A_{R\ae}(\theta,\gamma,m)$ for selected $\gamma$ and $m$.
The variations of the function with the angle $\theta$ are moderate,
which reflects the hidden symmetry between the PPV variables. 
}
\label{fig:3}
\end{figure}

The physical meaning of the new variable $\ae$ defined by 
Eq.~(\ref{eq:vzmap}) is that it defines 
the scale at which the turbulent velocity dispersion is
equal to $v$. In our turbulent model (see \S 3.1)  
we expect that after the transformation Eq.~(\ref{eq:vzmap}),
the asymptotic statistics of 3D PPV cubes will be approximately isotropic
and the three dimensional PPV distance $r_{\mathrm{PPV}}^2 = R^2 + \ae^2$
can be introduced.
More exactly, applying the transformation to Eq.~(\ref{ksicloud})
for the correlation function
\begin{eqnarray}
d_s(R,\ae) &=& 2\left(\xi(0,0)-\xi_s(R,\ae)\right) \nonumber \\
&\propto& \int_{-S}^S dz
\left[\frac{1}{z^{\gamma+m/2}} -
\frac{(R^2+z^2)^{1-\gamma/2-m/4}}{(1+m/2)R^2+z^2}
\exp\left(-\frac{F(\gamma,m)}{2}
\frac{\ae^m}{(R^2+z^2)^{m/2}}\right)\right]
\label{eq:dPPVraw}
\end{eqnarray}
and introducing the angle $\cos \theta=\ae/r_{\mathrm{PPV}}$,
$\sin\theta=R/r_{\mathrm{PPV}}$,
we see that the $r_{\mathrm{PPV}}$ scaling factorizes
\begin{equation}
d_s(R,\ae) \propto 
\left(r_{\mathrm{PPV}}\right)^{1-\gamma-m/2} A_{R\ae}(\theta,\gamma,m)
~~,~~~ r_{\mathrm{PPV}} \ll S
\label{eq:dPPV}
\end{equation}
when the integration is extended to infinity, as is appropriate for
$r_{\mathrm{PPV}} \ll S$. The integral form for the function
$A_{R\ae}(\theta,\gamma,m)$ that describes the angular dependence in $(R,\ae)$
space is given in Appendix~\ref{App:A}. This function is defined so that
$A_{R\ae}(\pi/2,\gamma,m)=A_{R0}(\gamma,m)$. With our choice of the 
coefficient $F(\gamma,m)$ in the transformation relation,
$A_{R\ae}(0,\gamma,m)=A_{R\ae}(\pi/2,\gamma,m)$.
Figure~\ref{fig:3} demonstrates that the variation of $A_{R\ae}(\theta)$
at intermediate angles is typically within ten percent.

In this discussion we have ignored the complication that arises from the
saturation of the structure function in the velocity direction. Indeed,
the line-of-sight asymptotics of Eq.~(\ref{AppCeq:d_rhoVf1}) is valid only for
$1-\gamma-3m/2 < 0$, a range much narrower than the
$1-\gamma-m/2 < 2$ allowed for Eq.~(\ref{AppC:eq_drhoR}). 
Analysis of this saturation involves taking the limit
of the integration in Eq.~(\ref{eq:dPPVraw})
in the way it is done in Section~\ref{sec:losdv}. 

The issue of the limited validity of structure functions
is resolved, if the power spectrum
is used directly. For this purpose
 the spectral formalism of LP00 proves to be useful. 
In Appendix~D we remind the reader of the
asymptotic results for the PPV spectrum $P(K,k_v)$ that
were obtained in LP00 using a velocity wave number $k_v$, reciprocal to 
$v/D_z^{1/2}(S)$. In these variables, the PPV spectrum is manifestly
anisotropic, having seemingly very different scalings along and perpendicular
to the line-of-sight. The hidden symmetries are revealed if  
$k_{\ae}\sim 1/\ae$ is used. 

The symmetry transformation of PPV statistical descriptors
allows one to introduce a new technique for
determining the turbulent velocity statistics from PPV cubes.
Since the mapping Eq.~(\ref{eq:vzmap}) depends explicitly on
velocity scaling $m$ and the amplitude of the turbulent motions $D_z(S)$,
one can fit for these parameters under the requirement that 
the structure function or the spectrum of the mapped PPV cube
acquire the same scaling in all directions. However this issue is
beyond the scope of the present paper. Within this paper the symmetries
allow us to generalize our asymptotics obtained for particular PPV directions
over a wider PPV volume.

\section{VCS for a Transparent Medium: $\alpha \rightarrow 0$}
The intensity in optically thin lines provides direct information
on the density in PPV space.
In this case, $\alpha \to 0$, the intensity is given by
the linear term in the expansion of the exponent in Eq.~(\ref{simplified})
\begin{equation}
I_v({\bf X})=\epsilon \rho_s({\bf X},v)~~.
\end{equation}
We shall treat separately several limits. First of all, in the limit
of high angular resolution (the narrow beam)
\begin{equation}
{\cal D}_{nar}(v) \propto \tilde d_s(0,v)
\label{narrow}
\end{equation}
while in the limit of poor angular resolution (the wide beam)
when the spectral data is effectively integrated
over the whole image of the object
\begin{equation}
{\cal D}_{wide}(v) \propto \int dR R \tilde d_s(R,v)~~~.
\end{equation}
We shall provide a criterion for the transition from one 
regime to another.

\subsection{High Resolution: Narrow Beam}

As the first step we consider the case of high resolution. The expressions
for $\tilde d_s (0,v)$ were discussed in \S 2.5, where the pitfalls related 
to the
direct use of structure functions along the velocity coordinate were 
demonstrated. In particular, it was shown  that the $v^2$ contribution arises
from the boundary effects. Here we analyze why this contribution is not
dominant for the VCS. 

The power spectrum is defined for narrow beam studies as 
\begin{equation}
P_{nar}(k_v) = -\frac{1}{\sqrt{2 \pi}}
\int_{-\infty}^{\infty} dv \cos[k_v v] {\cal D}_{nar}(v).
\end{equation}
Using Eqs. (\ref{eq:d_vV}), (\ref{narrow}) and (\ref{eq:rho_v_split}) 
we obtain
\begin{equation}
P_{nar}(k_v) \propto e^{-\beta k_v^2}
\int_{-1}^{1} {\mathrm d}\hat z\; |\hat z|^{-\gamma}
\exp\left[-\frac{1}{2} k_v^2 D_z(S) |\hat z|^m\right]~~~,
\end{equation} 
where as before $\hat z\equiv z/S$.
In the case where the integration over $\hat z$ is extended to infinity,
we just get
\begin{equation}
P_{nar,inf}(k_v) \propto \frac{2^{\frac{1-\gamma}{m}}}{m}
\Gamma\left[\frac{1-\gamma}{m}\right] 
e^{-\beta k_v^2}
\left[k_v D_z^{1/2}(S)\right]^{-2(1-\gamma)/m}~,  
\label{eq:Pn}
\end{equation} 
which coincides with the result obtained in the last Appendix of LP00
\footnote{Direct Fourier transform of the power-law
${\cal D}_{nar}= -2^{-p} m^{-1} \Gamma[-p] v^{2p}$ gives, of course,
the same result.}. There the same result was obtained using the
three-dimensional PPV power spectrum $P({\bf K},k_v)$ in PPV (see
Appendix~\ref{App:D}). Indeed, the
narrow-beam $P_{nar}(k_v)$ is given by the integral over ${\bf K}$
$P_{nar}(k_v)\propto \int d {\bf K} P({\bf K},k_v)$, thus,
 $P_{nar}(k_v) \propto P_1(k_v)$, where $P_1(k_v)$ is the 
one dimensional spectrum defined in LP00.

When the boundary effects are retained
\begin{equation}
P_{nar}(k_v) \propto 
\frac{2^{\frac{1-\gamma}{m}}}{m}
\left(\Gamma\left[\frac{1-\gamma}{m}\right]
-\Gamma\left[\frac{1-\gamma}{m},\frac{k_v D_z^{1/2}(S)}{2}\right]\right) 
e^{-\beta k_v^2}
\left[k_v D_z^{1/2}(S)\right]^{-2(1-\gamma)/m}~.
\end{equation}
Thus, the boundary effects in Fourier space
are localized to small k modes, through the incomplete Gamma function
term. At high $k_v \to \infty$ this term becomes negligible and we restore
the correct power law asymptotics even for very steep slopes $2/m > 3$
(a power spectrum slope of $-3$ corresponds to a structure function slope
$-2$ in 1D).

\subsection{Poor Resolution: Wide Beam}

The correlations of $I(v)$ are modified
when the instrument resolution is poor. In the limit
when the turbulent scale (or the whole cloud) is within the beam,
we effectively integrate PPV fluctuations over the image, i.e $R$ coordinate.
Namely,
\begin{equation}
{\cal D}_{wide}(v) \propto \int_0^\infty dR R \tilde d_s(R,v)
\end{equation}
Ignoring the boundary effects and angular dependence of $D_z({\bf r})$
(these do not affect asymptotic scaling) 
we can join the $R$ and $z$ integrations into a
three-dimensional integral to obtain 
\begin{equation}
{\cal D}_{wide}(v) \propto 
\frac{{\bar \rho}^2 S^2}{ D_z(S)}
\int  {\mathrm d}\hat r \; {\hat r}^2
\frac{1}{{\hat r}^{\gamma+m/2}} \left[ 1 -
\exp\left(-\frac{{\hat v}^2}{2 {\hat r}^m}\right) \right] 
\propto {\hat v}^{2 \frac{3-\gamma}{m}-1}~~,
\end{equation}
where once more the dimensionless variables $\hat v=v/D_z^{1/2}(S)$, $\hat r = r/S$ were used. 
This slope is very steep, e.g., for the pure velocity term, 
$\gamma=0$, $6/m-1 > 2$ for all $m < 2$, and as we have learned in \S 2.5
the boundary effects
saturate the direct structure function
 measurements at the universal, non-informative\footnote{The 
situation is a bit better if the density dominates and $\gamma > 0$, but
still the parameter range of sensitivity of the structure function remains
limited to high $m > 2(1-\gamma/3)$.} slope value
$2$. 

One again we should use 
the power spectrum which allows the non-informative terms to
be weeded out. At high wave numbers $k_v$ 
the one dimensional VCS in the wide-beam approximation, $P_{wide}(k_v)$,
is equal to the three dimensional PPV power spectrum $P({\bf K},k_v)$
taken at ${\bf K}=0$,
\begin{equation}
P_{wide}(k_v) \propto P({\bf K}=0,k_v) ~~~. 
\end{equation}
Three dimensional PPV power spectrum has been obtained in LP00. We have
\begin{equation}
P_{wide}(k_v) \propto (r_0/S)^{\gamma}
e^{-\beta k_v^2}
\left(k_v D_z^{1/2}(S)\right)^{-2 (3-\gamma)/m},
\label{eq:Pw}
\end{equation}
where the amplitude of the density contribution is defined through
the correlation length $r_0$.  For velocity contribution, $\gamma=0$,
\begin{equation}
P_{wide}(k_v) \propto 
e^{-\beta k_v^2}
\left(k_v D_z^{1/2}(S)\right)^{-6/m}
\end{equation}
which provides the high $k_v$ asymptotics for the VCS in the poor resolution
regime. 

\subsection{Transition from High to Poor Resolution}

A realistic beam has a finite width, $\Delta B$. We remind the
reader that we deal with the case in which  
the emitting volume extend along the line of sight is much
smaller than the distance to the volume. As the result,
the angular extend of the beam
is straightforwardly related to the physical scales that we deal with.

Whether the narrow (Eq.~(\ref{eq:Pn})) or
wide beam (Eq.~(\ref{eq:Pw})) regime is applicable depends on $k_v$.
To the linear scale $\Delta B$ corresponds the velocity
scale 
\begin{equation}
V_{\Delta B} \equiv \sqrt{D(S) (\Delta B/S)^m}~~,
\end{equation}
equal to the magnitude of turbulent velocities at
the separation of a size $\Delta B$. It is not difficult to find that when
\begin{equation}
k_v^{-1} > V_{\Delta B} 
\label{eq:nw}
\end{equation}
the beam is narrow, while on shorter scales its width is important.

The rigorous derivation of this criterion is straightforward. Results in
\S 3.5 and Appendix D allow to describe the VCS as the resolution changes.
However, the simplified argument below, that are
based on the already derived formulas,
gives a more intuitive way to obtaining of the result. Indeed, 
it is evident, that 
the difference between the narrow-beam and the wide beam power spectra is the 
integration over ${\bf K}$. One can approximate 
$P_{nar}(k_v) \propto (\Delta K)^2 P(0,k_v) = (\Delta K)^2 P_{wide}(k_v)$,
where $\Delta K$ is the size of Fourier domain over which the integral
is accumulated.  Comparing our results 
in Eq.~(\ref{eq:Pn}) and Eq.~(\ref{eq:Pw})
for the ideally narrow and fully-integrated beams, 
we find that $\Delta K \sim k_v^{2/m}$. The beam is narrow, if $\Delta B$
corresponds to Fourier space integration of at least the size $\Delta K$
and is wide otherwise. With proper dimensional coefficients taken into account,
we arrive to the criterion in Eq.~(\ref{eq:nw}). Another conclusion is that
the beam is effectively infinitely wide for all the scales of VCS study, if
the beam width exceed the scale at which the underlying structure function
of the turbulent velocity saturates (this scale is identified with S in this
paper).

\subsection{Expected Regimes for VCS}

\begin{figure}[ht]
\begin{center}
 \includegraphics[width=0.3\textwidth]{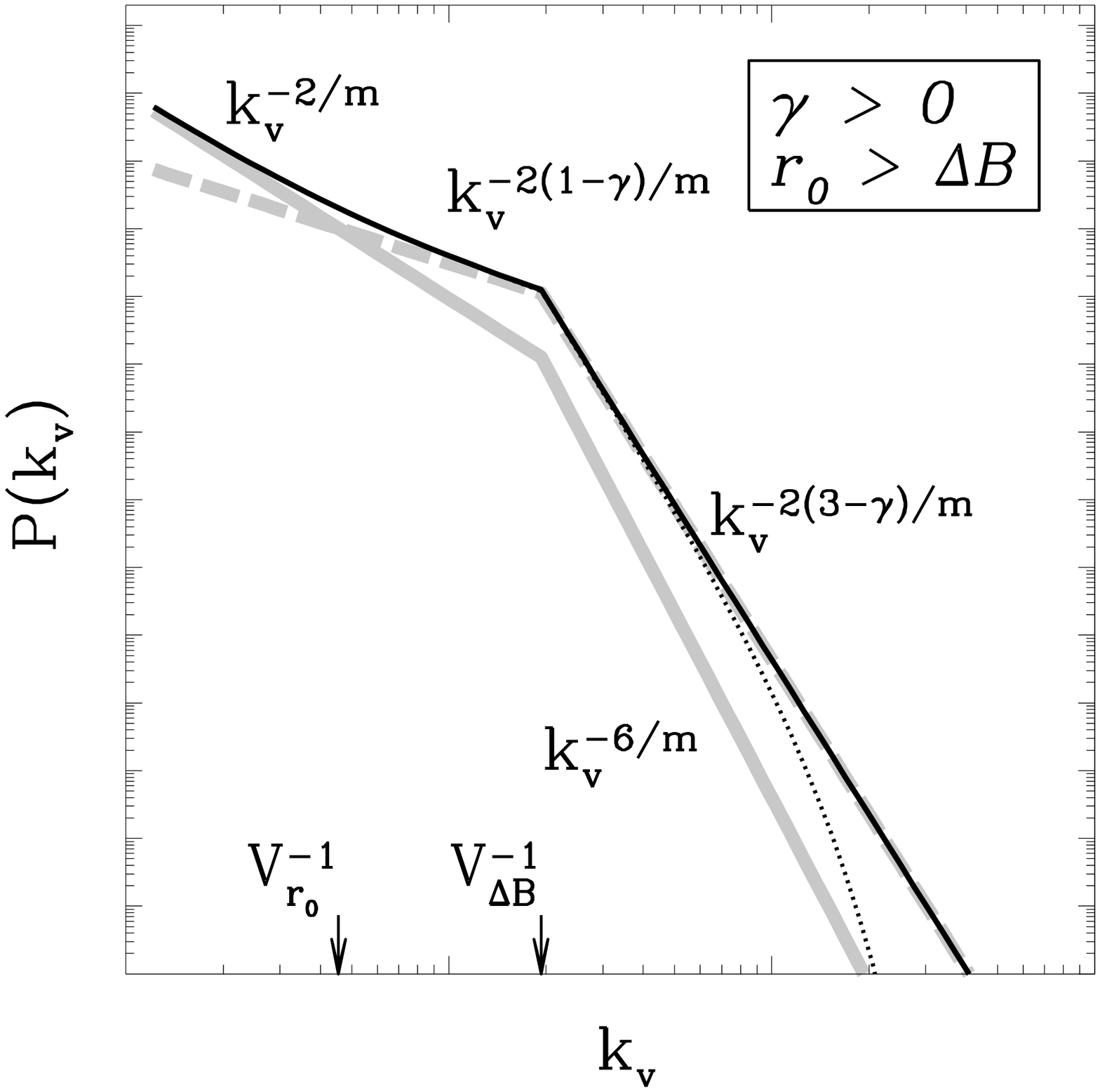} \hspace{3mm}
 \includegraphics[width=0.3\textwidth]{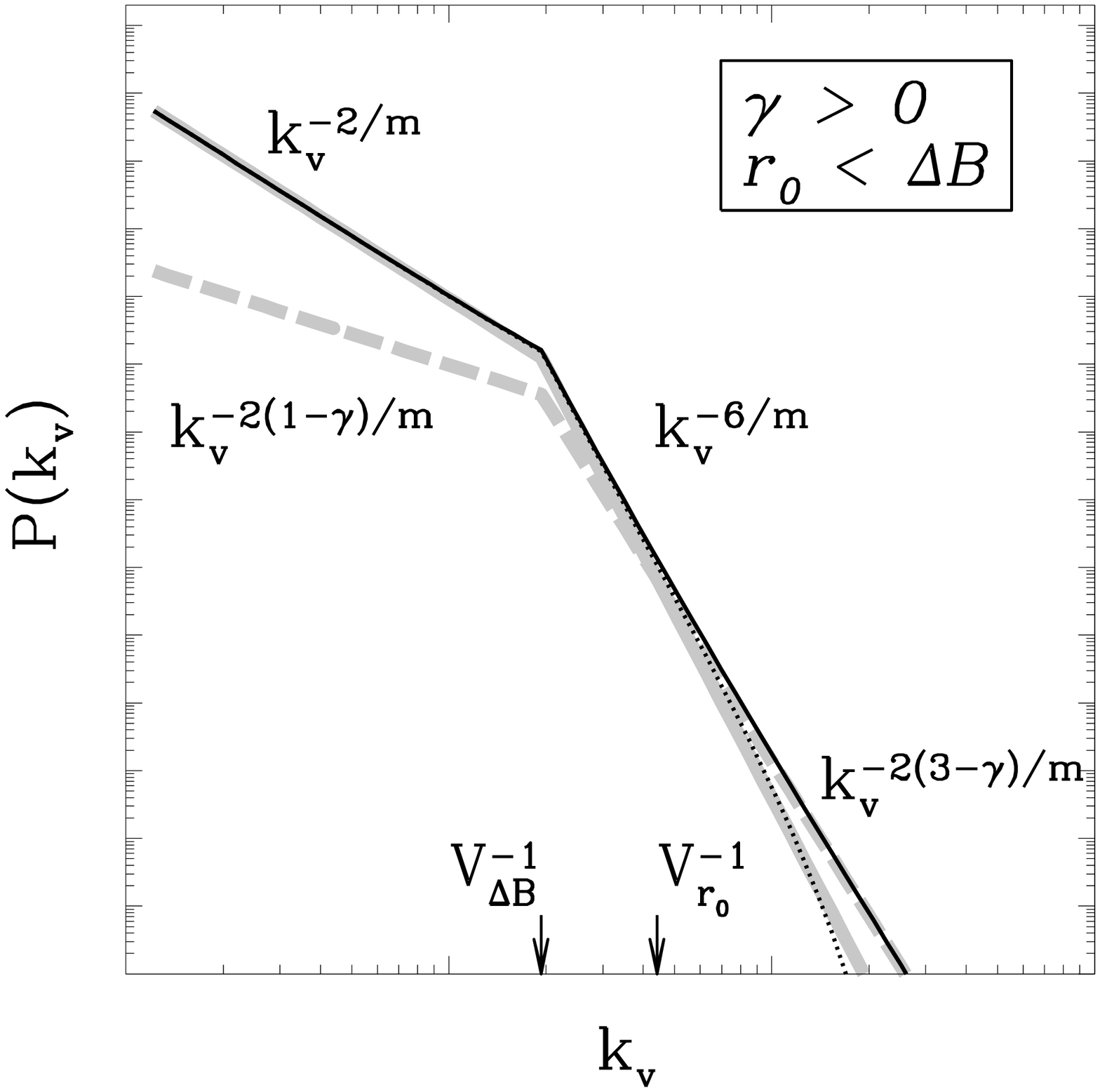} \hspace{3mm}
 \includegraphics[width=0.3\textwidth]{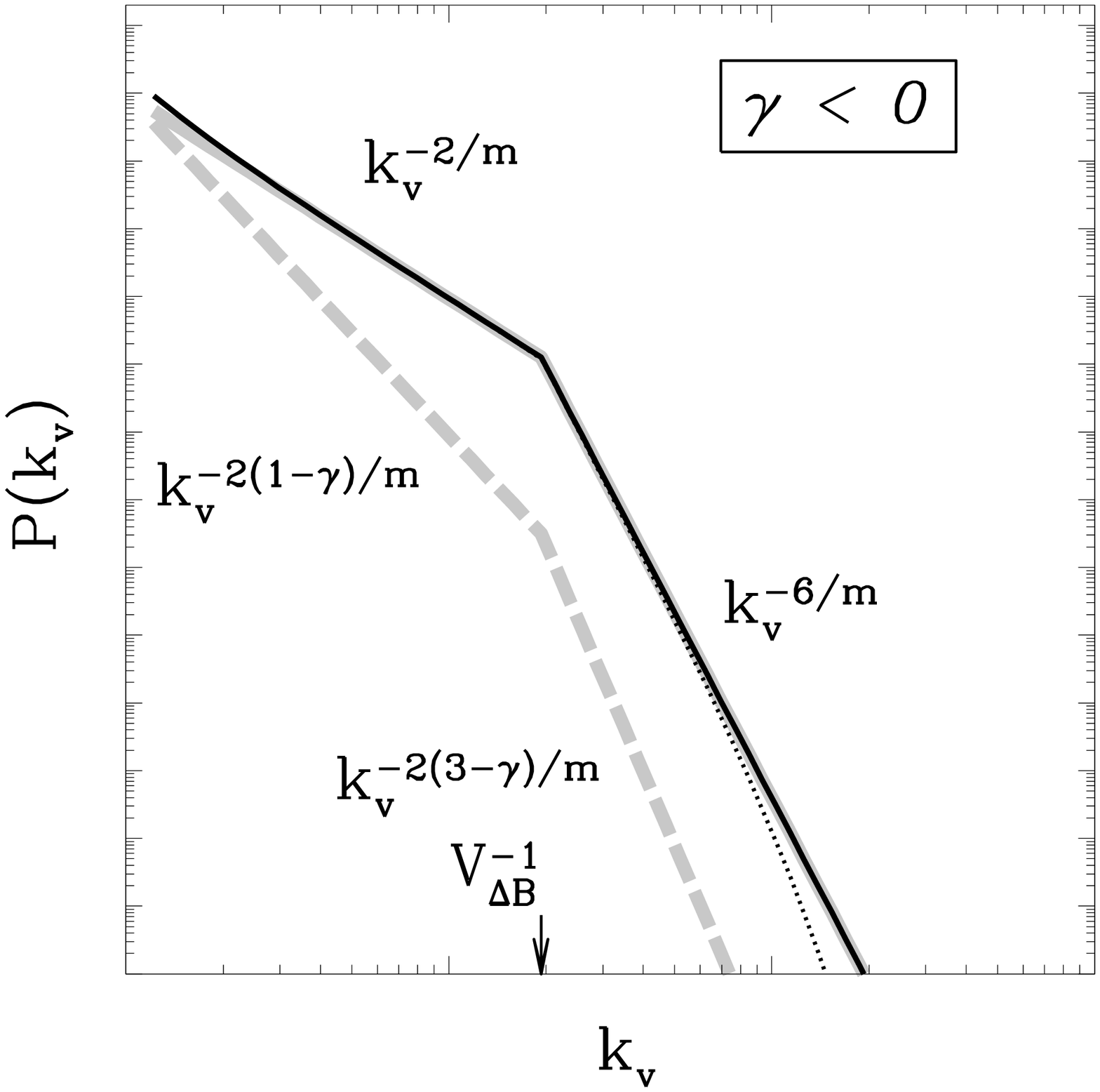} \\
\end{center}
\caption{ Qualitative representation of the density and velocity contributions
to the VCS power spectrum and the resulting scaling regimes.
In every panel light lines show contributions from the $\rho$-term
(density modified by velocity, dashed line) and $v$-term (pure velocity
effect, solid line) separately, while the dark solid line shows the combined
total VCS power spectrum. Thermal suppression of fluctuations
is shown by the dotted line. The labels above the dark solid curve
are arranged so as to illustrate
the sequential transition of the scalings of the total power spectrum.
Everywhere except the intermediate regimes, the total spectrum is dominated by
one of the components to which the current labeled scaling corresponds.
Labels below the dark solid lines mark the scaling of the subdominant
contributions.
For the left and middle panels the density
power spectrum is taken to be shallow, $\gamma > 0$.
The left panel corresponds to high amplitude of the density correlations,
$r_0 > \Delta B$, where density effects become dominant at relatively
long wavelengths for which the beam is narrow. In the middle panel,
the amplitude of density correlations is low $r_0 < \Delta B$ and they
dominate only the smallest scales which results in the intermediate steepening
of the VCS scaling. The right panel corresponds to the  steep density spectrum.
In this case the density contribution is always subdominant.
In this example the thermal scale is five times shorter than
the resolution scale $V_{\Delta B}$.
}
\label{fig:1}
\end{figure}

In Figure~\ref{fig:1} we summarize the different scalings of VCS.
As in our earlier papers (LP00, LP04) the main difference stems from
the density being either shallow or steep. If the density is shallow
i.e. scales as $\xi\sim r^{-\gamma}$, $\gamma>0$, which means
that the correlations increase with the decrease of the scale,
then it eventually becomes important at sufficiently small velocity
differences, i.e. at sufficiently large $k_v$. In the opposite case, i.e. when
$\gamma<0$, the contributions of density can be important only at
large velocity separations. 

The amplitude of the density
contribution to VCS is encoded in the correlation length $r_0$.
The velocity scale that corresponds to $r_0$ is
\begin{equation}
V_{r_0} = \sqrt{D_z(S) \left(r_0/S\right)^m}~~~.
\end{equation}
The comparison of the density ($\gamma$ is present)
and velocity ($\gamma=0$) contributions to VCS, given by Eq.~(\ref{eq:Pn})
or Eq.~(\ref{eq:Pw}) respectively, gives the critical scale
when they are equal
\begin{equation}
k_0^{-1} = C(\gamma,m) V_{r_0}
\end{equation}
The numerical factor 
$C(\gamma,m)=\sqrt{2}
\left(\Gamma[1/m]/\Gamma[(1-\gamma)/m]\right)^{\frac{m}{2\gamma}}$
for a narrow beam
and is somewhat different for a wide beam,
but in both cases it is of order of unity in the interesting range $-1 < \gamma
< 1 $ (several aspects of our formalism break down at $\gamma = 1$).
So we can roughly equate the scale of equality of the density and velocity
effects to $r_0$, i.e., in velocity units, 
\begin{equation}
k_0^{-1} \approx V_{r_0} ~~.
\end{equation}

The left and middle panels of Figure~\ref{fig:1} deal with the case of shallow density. 
Velocity is dominant at $k_v < k_0$, while the density term provides the 
main contribution at $k_v > k_0$. The left panel demonstrates the case where
the scale of transition from asymptotics is entirely
dominated by velocity to the one influenced by spatially resolved velocity and density, $V_{\Delta B} < V_{r_0}$.
The observed fluctuations arising from
 the unresolved turbulent eddies depends on the scalings of both
velocity and density. In the middle panel, $ V_{\Delta B} > V_{r_0}$,
the transition scale is unresolved. In this case
if there is still a dynamical range for
moderately long scales to be resolved by the experiment
$D_z(S)^{1/2} > k_v^{-1} > V_{\Delta B}$, 
the VCS of the resolved eddies will be determined by the turbulent velocities
only.

The right panel of Figure~\ref{fig:1} addresses the case of a steep density
spectrum. The difference now is that fluctuations of the density are maximal
at low wave-numbers and it is there that the density
could be important.  Velocity is dominant at shorter scales $k_v > k_0$.
However, as we discussed in \S 2.3, 
the steep density correlation length $r_0$ is large,
at least as large as the density power cutoff $r_c$,
which argues for density fluctuations to be
subdominant everywhere up to the scale of the emitting turbulent volume
(``cloud''), which is the range of scales that we consider here.

\subsection{Thermal Broadening and VCS Cookbook}

\subsubsection{Thermal effects and inertial range}

In Figure~\ref{fig:1} we have plotted the power spectra over 3 decades of
velocity magnitude to compactly demonstrate all possible scalings.
We clearly see two contributions to the VCS, one
part arising from
pure velocity effects in uniformly distributed matter, while the 
other represents the contribution of density
inhomogeneities modified by the velocity. Thermal effects are shown as well.

In observations, one can  anticipate a coverage over
two decades of velocity magnitudes before thermal effects
get important. Potentially, correcting for the thermal
prefactor $\exp(-\beta k_v^2)$ in Eqns.~(\ref{eq:Pn},\ref{eq:Pw}),
one can extend the observational range further. Note, that the thermal
corrections are different for species of different mass. Therefore, by using
heavier species one can extend the high $k$ cut-off by the square root of
the ratio of the mass of the species to the mass of hydrogen. 

How small subsonic (in Cold gas) scales can still be probed depends
on the signal-to-noise ratio of the available data. Indeed, 
the ability to deconvolve the thermal smoothing is limited by noise
amplification in the process.
However, any extension of the VCS by a factor $a$  in velocity
results in the extension of the sampled spatial scales by a factor of
$a^{2/m}$, which is $a^3$ for the Kolmogorov turbulence.  
   
Even with a limited $k_v$ coverage, important results can be obtained
with VCS, especially if observations encompass one of
the transitional regimes.
For example, if one measures a transition from
a shallow spectrum to a steeper one (see the left and the right panels
in Figure~\ref{fig:1})
one has the potential to i) determine the velocity index $m$,
since the difference between the slopes is always $4/m$;
ii) determine $\gamma$ next;
iii) estimate the amplitude of turbulent velocities from the position
of the transition point as discussed above.
On the other hand, if one encounters a transition from a steep to a shallower
spectrum, one i) may argue for the presence of the shallow density
inhomogeneities;
ii) estimate $\gamma/m$ from the difference of the slopes, and then $m$;
iii) estimate the density correlation radius $r_0$. Finally, if no transition
regime is available, as is the case of strong absorption (see 
Figure~\ref{fig:2}) than when $\gamma <0$ one can get $m$, while for the case
of $\gamma>0$ a combination of $\gamma$ and $m$ is available.

\subsubsection{Separating thermal and non-thermal velocities}

Consider HI in the Warm and Cold phases as an example. If one disregards,
for the sake of simplicity, the cross-correlation between the fluctuations
in these two phases, then the VCS
from such a system will be a sum of the spectra from $T_{cold}\sim
100$~K and
$T_{warm}\sim 10^4$~K gas. It is easy to see that
the contribution of the Warm gas to the total VCS may be neglected
for all velocities that are subsonic for Warm gas, i.e.
for velocities less than $10$ km/s.
Because the suppression of the Warm gas contribution is exponential,
the VCS would reflect the turbulence in the
Cold gas, even if the mass fraction of the Cold gas is small.
Therefore, to correct for the thermal velocities, one should
multiply\footnote{In a similar way one can treat the velocity resolution
of the instrument. However, in this case, similar to the case of the
thermal broadening corrections, the extent to which the VCS can be extended
is limited by the noise amplification that the procedure entails.}
the VCS power spectrum by $\exp(\beta_{cold} k_v^2)$.

In a more complex model with HI at intermediate temperatures (Heiles \&
Troland 2003) the cold gas still dominates the VCS.  In fact, 
the machinery that we have developed here
allows us to make an
independent test of the HI temperatures measured by other techniques. The
ability of the VCS to separate of thermal and non-thermal contributions
to the line-width makes it a unique tool.
One can apply spatial smoothing to the PPV data of a cloud.
As the result, the position
of the knee between the high and low resolution spectra
(see Figure~\ref{fig:1}) will change. The inverse
wavenumber at the knee $k_{knee}^{-1}$ corresponds to the non-thermal
velocity at the scale of smoothing $\theta_{smooth}$. Thus we can establish
the amplitude of the non-thermal velocities, in particular at the
scale of the whole cloud 
$D_z^{1/2}(S) \approx k_{knee}^{-1} (\theta_{cloud}/\theta_{smooth})^{m/2}$,
where $m$ is the index of the velocity structure function that is being
determined by the VCS (this index is $m=2/3$ for Kolmogorov turbulence)
and $\theta$'s are angular sizes of the cloud and the smoothing length.
This allows one to measure
cloud gas temperatures. We shall discuss the details of this
technique elsewhere. Choosing the correct factor of $\exp(\beta_{cold} k_v^2)$
to straighten the VCS plot can be another procedure for determining the
cold gas temperature (e.g. Chepurnov \& Lazarian 2006).

\subsubsection{VCS cookbook}

The VCS cookbook  is rather straightforward. VCS
near a scale $k_v$ depends on whether the instrument  
resolves the correspondent spatial scale
$\left[k_v^2 D_z(S)\right]^{-1/m} S$,
where $S$ is the scale where turbulence saturates. 
If this scale is resolved then $P_v(k_v) \propto k_v^{-2/m}$
and $P_{\rho}(k_v) \propto k_v^{-2(1-\gamma)/m}$. 
If the scale is not resolved then
$P_v(k_v) \propto k_v^{-6/m}$ and $P_\rho(k_v) \propto k_v^{-2(3-\gamma)/m}$. 
These results are presented in a compact form in Table~\ref{table:results}.
\begin{table}[h]
\centering
\begin{tabular}{lll} \hline\hline \\[-2mm]
Spectral~term & $\Delta B < S \left[k_v^2 D_z(S)\right]^{-\frac{1}{m}}$ &
$ \Delta B > S \left[k_v^2 D_z(S)\right]^{-\frac{1}{m}}$ \\[2mm]
\hline \\[-2mm]
$ P_\rho(k_v) $ & $ \propto\left(k_v D_z^{1/2}(S)\right)^{-2(1-\gamma)/m} $& 
$ \propto\left(k_v D_z^{1/2}(S)\right)^{-2(3-\gamma)/m} $ \\[2mm]
\hline \\[-2mm]
$ P_v(k_v) $ & $ \propto\left(k_v D_z^{1/2}(S)\right)^{-2/m} $  & 
$ \propto\left(k_v D_z^{1/2}(S)\right)^{-6/m} $ \\[2mm]
\hline
\end{tabular}
\caption{Scalings of VCS for shallow and steep densities.}
\label{table:results}
\end{table}
The transition from the low to the high resolution regimes happens as
the velocity scale under study gets comparable to the turbulent velocity
at the minimal spatially resolved scale. As the change of slope is the
velocity-induced effect, it is not surprising that the difference in
spectral indexes in the low and high resolution limit is $4/m$ for both
$P_v$ and $P_\rho$ terms, i.e it does not depend on the density\footnote{In
the situation where the available telescope resolution is not sufficient,
i.e. in the case of extragalactic turbulence research, the high spatial
resolution VCS can be obtained via studies of the absorption lines from
point sources.}.
This allows for separation of the velocity and density contributions.
For instance, Figure~\ref{fig:1}
illustrates that in the case of shallow density both the density and velocity 
spectra can be obtained.

Note that obtaining the density spectrum from a well resolved
map of intensities is trivial for the optically thin medium,
as the density spectrum is directly available from the column densities (i.e.
velocity integrated intensities). However, for the absorbing
medium such velocity-integrated 
maps provide the universal spectrum $K^{-3}$,
where $K$ is the 2D wavenumber (LP04). Similarly, even for the optically thin
medium, it is not possible to get the density spectrum if the turbulent
volume is not spatially resolved. On the contrary, $P_{\rho}(k_v)$ reflects
the contribution of shallow density even in this case (see Figure~\ref{fig:1}).

\section{Effects of Absorption}

The main {\it scale dependent} effect of absorption is to
diminish correlation between intensity of emission at
widely separated velocities or lines of sight, i.e from distant points
in PPV space. The effect of absorption has been discussed at length
in LP04 in the framework of the VCA technique. Here we capitalize on the
insight obtained there.

Along the velocity coordinate the separations that are affected by absorption
can be estimated using Eq~(\ref{dv_PPVidexp}). Similar to LP04
we get that the absorption 
becomes important when $\alpha^2 d_s(0,v) > 1$,
thus for $v$ exceeding
the absorption velocity scale $V_{ab}$ which we define as
\begin{equation}
\alpha^2 \tilde d_s(0,V_{ab}) =1~~~~~~~.
\label{vab}
\end{equation}

Let us consider in detail the $\gamma < 0$ case when the velocity term
dominates VCS.  Asymptotic expressions for $d_v$  given by
Eq.~(\ref{eq:d_vVfg23}) and (\ref{eq:d_vVfl23})
lead to the absorption window width
\begin{eqnarray}
V_{ab}/D_z(S)^{1/2}
 &\approx& \left(\alpha \bar \rho_s\right)  ^{\frac{2m}{m-2(1-\gamma)}}, 
~~~~~~ m > 2/3 \label{eq:abs_width1}\\
V_{ab}/D_z(S)^{1/2}
 &\approx& \left(\alpha \bar \rho_s\right)^{-1} ~~~~~~~,
~~~~~~ m < 2/3
\label{eq:abs_width2}
\end{eqnarray}
where we have omitted numerical coefficients of order unity and have estimated
$\bar \rho_s = \bar \rho S /D_z(S)^{1/2}$.

For our purpose of studying turbulence over its power-law
inertial range, the only velocity range that matters is $v < V_{ab}$. 

When the resolution of the instrument is finite and signal is averaged
over an angular area, absorption determines
 how distant lines-of-sight contribute
to the correlated signal. This effect is described by the absorption window
$W_{ab}(R)$, introduced in \S 2 (see Eq.~(\ref{Wabs})), which behaves similar to the instrument
beam, downweighting the distant pairs.

Let us estimate the form of the window assuming Gaussian statistics
and homogeneity of the density fluctuations
$\delta \rho_s = \rho_s - \langle \rho_s \rangle $. Then (see LP04)
\begin{equation}
W_{absorption} \approx \langle e^{-\alpha \rho_s({\bf X_1},v_1)} \rangle
\langle e^{-\alpha \rho_s({\bf X_2},v_1) } \rangle
e^{\frac{\alpha^2}{2} \langle \delta\rho_s^2({\bf X_1},v_1)+ 
\delta\rho_s^2({\bf X_2},v_1) \rangle}
e^{-\frac{\alpha^2}{2}{\tilde d}_s(R,0)}
\label{Wabs_gaussian_final}
\end{equation}
The last term is the one that determines the scale dependence of the window.
The most important qualitative
characteristic of the window is its width, which for absorption we shall
define as $R_{ab}$ at which 
\begin{equation}
\alpha^2 \tilde d_s(R_{ab},0) =1~~~~~~~.
\label{Rab}
\end{equation}
In terms of the velocity-density decomposition of the PPV structure function,
the product of the two windows arises, which is $\sim e^{-\alpha^2 \tilde
  d_v(R,0)/2} e^{-\alpha^2\tilde d_\rho(R,0)/2}$. Both factors act
simultaneously but the one with the smallest width determines the gross
effect. This is $d_v(R,0)$ for $\gamma < 0$.
Setting $\gamma=0$ in Eq.~(\ref{AppC:eq_drhoR}) to obtain the asymptotics
for $d_v(R,0)$ we get
\begin{equation}
R_{ab}/S \approx 
\left(\alpha \bar \rho_s\right)  ^{\frac{2}{m-2}}  ~~~.
\end{equation}

One must compare the absorption window scale $R_{ab}$ to the beam $\Delta B$
on one hand, and to the scales under study $v$ on the other hand.
When $R_{ab} > \Delta B$, the absorption window is not important.
In the opposite regime, $R_{ab} < \Delta B$, 
when the absorption affects the angle
averaging, one compares the velocity scale, corresponding to $R_{ab}$,
$V_{ab}=D_z^{1/2}(S) \left(R_{ab}/S\right)^{m/2}$, with $v$.
For $v < V_{ab}$ the
effective window is a low resolution one, while for $v > V_{ab}$ absorption
seems to induce an effective high resolution beam.
Given the symmetry between
the line-of-sight and orthogonal sky correlations that we discussed in \S 3.5,
it is, however, no surprise that $V_{ab}$ defined this way coincides with the 
critical velocity separation given by Eq.~(\ref{eq:abs_width2}) above which
absorption destroys the power-law scaling solutions. Hence, one cannot
recover the high resolution scaling laws with the help of absorption effects.

\begin{figure}[ht]
\begin{center}
 \includegraphics[width=0.3\textwidth]{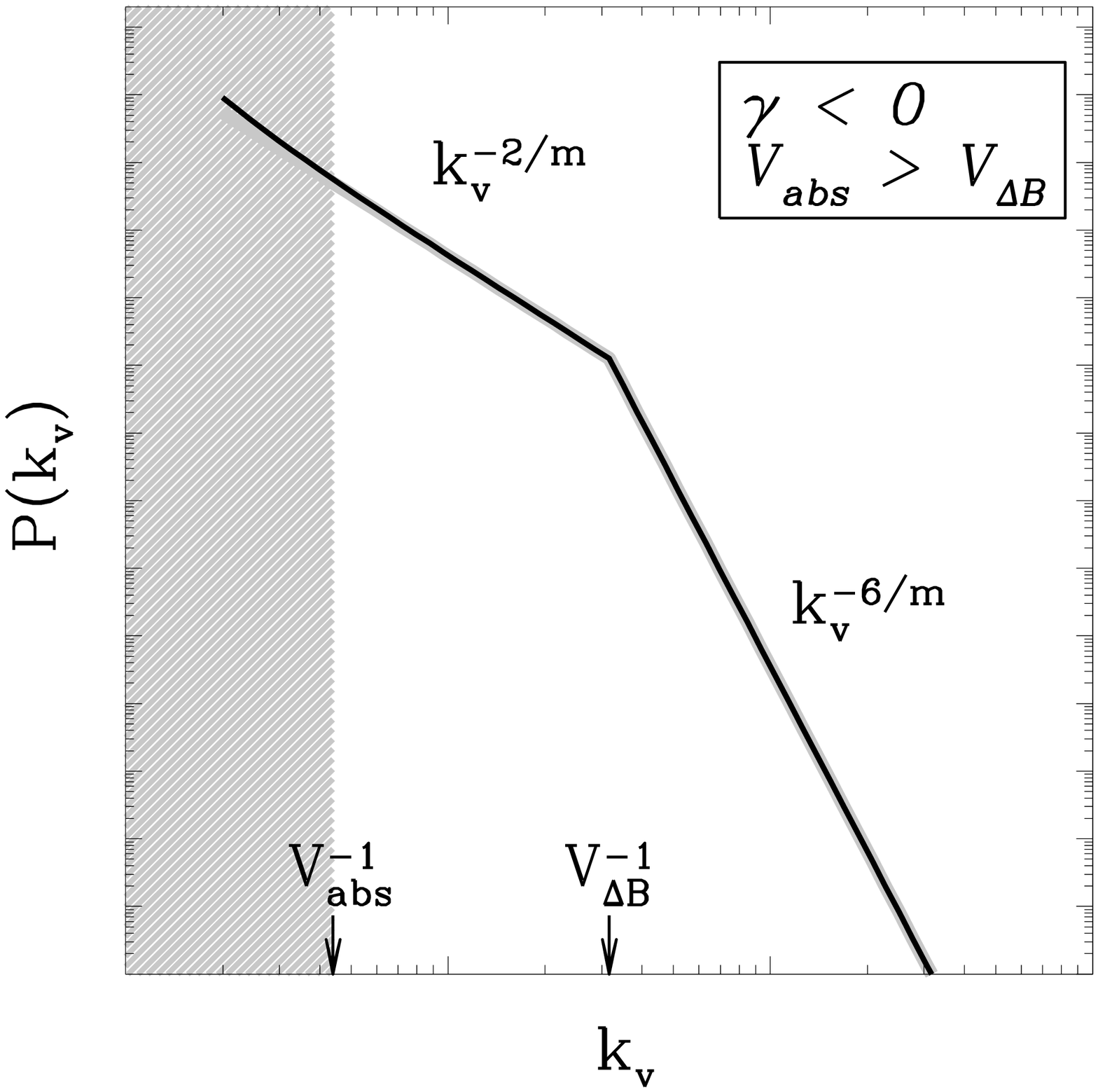} \hspace{3mm}
 \includegraphics[width=0.3\textwidth]{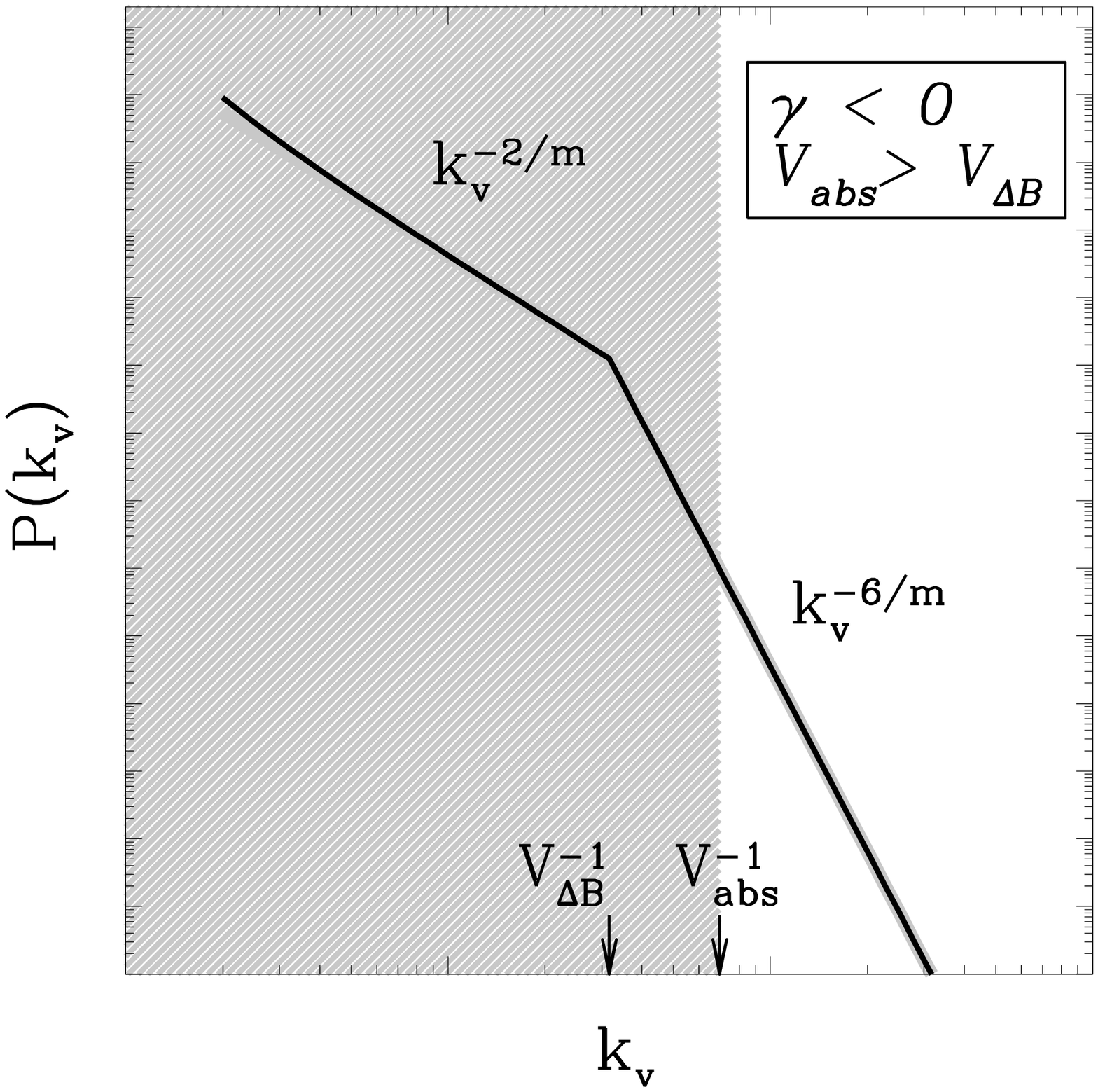} \\
\end{center}
\caption{ Qualitative representation of the absorption effects
on the VCS power spectrum and the resulting scaling regimes.
Shaded region corresponds to the regimes obscured in the presence of 
absorption.
}
\label{fig:2}
\end{figure}

Thus, we conclude that to study the inertial turbulence range one should focus
on $v < V_{ab}$ and to be able to measure high-resolution asymptotic VCS
one must have instrument resolution better than the 'absorption window,
$ V_{\Delta B} < V_{ab}$ (see Figure~\ref{fig:2}).

\section{Discussion}

\subsection{Simplifications Employed}

To simplify our presentation we considered emission that is proportional to 
the first power of density. However, we found that for many cases, e.g. for
steep density, the actual  spectrum of density is irrelevant. Therefore these 
results are not affected by the actual assumptions about the scaling of
 emissivity. In general, our results may be trivially generalized if 
correlation functions of emissivities are used instead of correlation 
functions of densities (Chepurnov \& Lazarian 2006).
With this in mind we may claim that the VCS is applicable not only to HI,
but also to CO transitions, emission by ions in 
turbulent plasmas, various molecules etc.

Following LP04 in the paper we employed a simplified
treatment of the radiative transfer. However, it is
 explained in
LP04 that a more elaborate treatment of absorption cannot change the 
high wavenumber asymptotics that we are interested in. In short, 
at low $k_v$ 
the absorption gets important and it destroys the power-law behavior.
We do not have predictions for the VCS in this regime.
However, our results show
that at sufficiently high $k_v$ (which depends on the window function
determined by the absorption) the power-laws are not affected and
the VCS can be used to study turbulence.

Note, that in the paper above we have analyzed the problems of using structure
functions along the velocity coordinate
 being very steep.
One possible way of proceeding with a real space statistical description
in this case is to design a 'next order' structure function
which will have a counter-term eliminating the contaminating
quadratic in velocity 
term.
Such 'next order' function will be insensitive not only to the constants
added to the field, but also to the gradients of the field.
We shall not develop higher order structure function formalism here. 
(Eq.~(\ref{eq:Gamma}) gives a good guidance on what terms need
 to be canceled). Instead, we argue for using spectra along the velocity
coordinate which allows us to avoid a lot of unnecessary trouble.

\subsection{VCS and Other Techniques}

\subsubsection{VCS \& VCA: optically thin lines}

A unique feature of the VCS is that this technique utilizes information that
has not been used before, as far as we know. Indeed, 
we developed the VCA in LP00 to explain the puzzling data on spectra of PPV 
slices (see Green 1993).
The motivation for LP04 was both to study the domain of applicability of 
VCA in the presence of absorption and to explain the CO line integrated data (see Stutzki et al. 1998).
The situation is different with the VCS, where the practical application of 
the technique to the actual data has just started (see Lazarian 2004,
2005, 2006, Chepurnov \& Lazarian 2006).   

The earlier techniques to study turbulence, e.g. velocity centroids or VCA
 require the observations to spatially resolve the scale
of the turbulence under study\footnote{As it was discussed in LP00, the VCA
can be applied directly to the raw interferometric data, rather than to
images that require good coverage of {\it all} spatial frequencies. 
However, even
with interferometers, the application of the VCA to extragalactic objects is
restricted.}. This constrains the variety of astrophysical
objects where the turbulence can be studied.
The technique presented here, namely, the VCS,
is a unique tool that allows studies of astrophysical turbulence even when
the instrument does not resolve the turbulent fluctuations spatially. Indeed, 
it is essential for the technique to resolve only the
fluctuations of intensity along the V-axis (see \S 4.2).
This is of immense importance for studies of turbulence in poorly
resolved extragalactic objects, supernova remnants and circumstellar
regions\footnote{For some of these objects the issue of isotropy and 
homogeneity of turbulence may arise. Indeed, if the properties of turbulence
change substantially along the line of sight, the measured spectra would 
represent the averaged properties of turbulence. To get more detailed 
information one may need to average the fluctuation arising, for instance,
from a supernova remnant, over annuli over the image, which would require
some spatial resolution, but may not still require resolving the spatial 
turbulent scale under study.}. 
 There is also potential for the modification of the technique for
laboratory research, e.g. for plasma turbulence studies.

Our study of the effect of finite temperatures for the technique reveals that,
unlike the VCA, the temperature broadening does not prevent the turbulence
spectrum from being recovered from observations. Indeed, in VCA, gas
temperature acts in the same way as the width of a channel. Within the VCS
the term with temperature gets factorized. One can correct for this term,
e.g. by fitting for the temperature that would remove the exponential
fall off in the spectrum. This also allows for a new
way of estimating the interstellar gas temperature 
(see Chepurnov \& Lazarian 2006).

Another advantage of the VCS compared to the VCA is that it reveals the
spectrum of turbulence directly, while within the VCA the slope of the spectrum
should be inferred from varying the thickness of the channel. As the thermal
line width acts in a similar way as the channel thickness, additional care
(see LP04) should be exercised not to confuse the channel that is still
thick due to thermal velocity broadening with the channel that shows the 
thin slice asymptotics. We feel that a simultaneous
use of VCA and the VCS makes the turbulence
spectrum identification more reliable.

\subsubsection{VCS \& VCA: self-absorption and absorption lines}

The introduction of absorption in VCS and VCA brings about different results.
Within the analysis of velocity slices spectra (VCA) the absorption results
in new scalings for slices for which absorption is important. 
The turbulence spectral indexes 
can be recovered for the VCA within 
sufficiently thin slices, provided that the thickness of the slices
exceeds the thermal line width. For the VCS no new power-law 
asymptotics is available in the presence of absorption. When absorption 
becomes important the spectra get exponentially damped. This
simplifies the interpretation of the
data.

As we discussed in LP00 and LP04 the VCA is applicable to studies of not only 
emission, but also absorption lines. An example of such a study is provided
in Deshpande, Dwarakanath \& Goss (2000), where an extended synchrotron emission background was used to map the absorption in the HI gas. The necessity of
using extended emission sources limits the extent of possible VCA studies of
turbulence. This is not an issue for the VCS, for which absorption 
lines\footnote{The mathematical formulation to the problem of the VCS for
absorption lines are very similar to the VCS for emission lines. The absorption
for optical and UV lines for which stimulated emission is negligible is 
proportional to $\int_0^S dz \rho ({\bf x}) \phi_v ({\bf x})$ and therefore 
measurements of the absorbed intensity can be treated the same way as the
intensity of the emission spectral line in 
\S 2.1. The difference is that this case is simpler as one should not worry 
either of the finite resolution or absorption effects. In the case of HI 
absorption study of the ratio of density over temperature
enters instead of density. In the case of an isobaric medium the product of
density and temperature are constant and the problem is similar to studies of 
transitions for which the emissivity is proportional to $\rho^2$. In general,
our study shows that the VCS in most cases is dominated by velocity 
fluctuations. Thus we expect that the temperatures should not much affect the
VCS even for HI absorption line studies.} from {\it point sources} can be
used. Interestingly enough, in this case the VCS asymptotics for the high
resolution limit should be used irrespectively of the actual beam size of the
instrument.

\subsubsection{Relation to other techniques}

 Talking about the role of the present study in a more general framework 
of the techniques for studies of turbulent velocities we would say that
this paper provides a formalism that
should allow to describe the properties of Spectral Correlation Functions 
(SCF) 
(see Padoan, Goodman \& Juvela 2003 and references therein) in the 
presence of absorption. The relation, however, between these techniques and another statistical tool, namely,
Principal Component Analysis (PCA) 
(see Heyer \& Brunt 2004 and references therein) is
not yet clear. 

Taking turbulence studies in a broader context, we may note, that
similar to the successful studies of electron density fluctuations
in the ionized media 
(see Spangler \& Gwinn 1990) the VCS can study
turbulence without requiring high spectral resolution.
 Note, however,  that the aforementioned 
 scintillation studies have limitations 
arising from limited number of sampling directions as well as from the
technique being relevant only for
ionized gas at extremely small scales.
Moreover, these sorts of measurements provide only the density statistics, 
which is an indirect measure of turbulence. Naturally, combining
information on turbulence obtained by different channels provides
big advantages for constraining the models of interstellar turbulence and
testing its correspondence with the theoretical expectations\footnote{For
instance, the issues of whether the Alfvenic turbulence has Kolmogorov spectrum
of $-5/3$ or the Iroshnikov-Kraichnan spectrum of $-3/2$ have been widely
discussed in the literature (e.g. Maron \& Goldreich 2001, Muller, Biskamp \&
Grappin 2003, Biskamp 2003, 
Boldyrev 2005, Beresnyak \& Lazarian 2006). The difference in the VCS slopes
for these two underlying spectra is $1$, which can be found through 
observations. Defining the type of spectrum is important for many
applications, including thermal conduction,
cosmic ray and cosmic dust dynamics (see Lazarian 2006ab, Cho \& Lazarian 2006,
Yan, Lazarian \& Draine 2004 and references therein.}.

\subsection{Progress and Prospects}

While establishing the mathematical foundations of the VCS, in
this paper we have improved the statistical description of the PPV.
In particular, we established the source of problems related to the use
of structure functions along the velocity coordinate and proposed a remedy.
In addition, we corrected our earlier statements about the relative
importance of density and velocity contribution in the case when the
density is steep. As the result, we established that the for a steep density
spectrum, the density contribution is always subdominant. This fact simplifies
practical studies of turbulence both using the VCS and the VCA. 

We have extended the general theory of the correlations in PPV space.
Potentially,
although we did not pursue this within this paper, our findings on the
symmetries existing in the PPV space open a way to study the velocity field
using the entire PPV data cubes, rather than slices, as we use in the VCA, or
along-the-line statistics as we use in the VCS. This also allows one to extend
the centroids and modified centroids (Lazarian \&
Esquivel 2003, Esquivel \& Lazarian 2005) techniques for studies of 
turbulence in the presence of absorption.

In \S 4.5 we have discussed the thermal broadening
 using HI as an example. 
Using heavier species that exhibit lower thermal broadening
allows one to study turbulence up to smaller scales.  One can extend the
range of environments that can be proved by the VCS using different wavelengths. For
instance, the X-ray spectrometers with high spatial resolution can be used
to study of turbulence in hot plasma. In particular, the potential of VCS is
high for studies of plasma turbulence in clusters of galaxies (see
Sunyaev et al. 2003 and references therein).  A simulated
example of such a study with the 
future mission Constellation X is provided in Lazarian (2006). 

Studies of turbulence in objects which are poorly resolved spatially is a
natural avenue for the VCS applications. Interestingly enough, in this case
one can combine the absorption line studies, which would provide the 
VCS for the high resolution, with the emission 
studies that would provide the VCS in the poor resolution limit. Potentially,
both velocity and density spectra can be obtained this way.

The importance of this work goes beyond the actual recovery of the
particular power-law indexes. First of all, the technique can be generalized
to solve the inverse problem to recover non-power law turbulence spectra.
This may be important for studying turbulence at scales at which
either injection or dissipation of energy happens\footnote{Consider turbulent
damping scale. The study can provide us with the angular scale at which
damping occurs, i.e. with the ratio of $d_{diss}/L$, where $L$ is the distance
to the cloud under study. Thus if we know the distance, we can have
an insight into the damping of turbulence and therefore the physical conditions
in the cloud. Alternatively, there are situations where the distance to the
cloud is notoriously poorly known, as this is the case of high velocity
clouds (see Wakker 2004). In such situations the calculations of $d_{diss}$ scale
(see expressions (20) and (6) in Lazarian et al. (2004), where $d_{diss}$ 
should be identified with the inverse of the critical perpendicular wavenumber)
can be provided with higher accuracy using the velocity dispersion at the
scale of the cloud. This allows to place better limits on $L$.}. Such studies are important
for identifying astrophysical sources and sinks of turbulent energy.
Second, studies of the transition from low resolution to high resolution
regimes allows one to separate thermal and non-thermal contributions
to the line-widths as has been discussed in \S 4.4. This could both
test the thermal correction that can be applied to extend
the power-law into sub-thermal velocity range (see \S 4.4 and Chepurnov
\& Lazarian, 2006) and enable studies of temperature distribution
of the gas in atomic clouds (Heiles \& Troland, 2003).

\section{Summary}

I. VCS is a new technique to study astrophysical turbulence
in interstellar gas, intracluster plasma, supernova remnants etc. Its
major advantage to the existing techniques is that it does not necessarily
require high spatial resolution to recover the information on turbulence.

II. VCS can employ both absorption and emission lines to study turbulence.
The current study concentrates mostly on emission lines and develops a
mathematical formalism that may be modified to deal with absorption lines.
We account for the effects of thermal broadening and self-absorption. 

III.  VCS has two parts,
one depending exclusively on velocity and the other depending on both velocity
and density: the relative amplitude of the two terms depends on the amplitude
of density perturbations and the dominance of the particular part depends
on whether density statistics is shallow or steep:

a) if the density statistics is steep ($P (k) \propto k^{-n} , n > 3$),
the VCS is affected only by the turbulent
velocity.

b) if the density spectrum is shallow, then at small wave-numbers the VCS
 is affected only by the velocity fluctuations
 and at larger
wave-numbers - by both density and velocity fluctuations. The wavenumber corresponding
to the transition point between the two regimes above depends on
the amplitude of the density fluctuations.

IV. The particular power-law indexes of the VCS depend on whether
the fluctuations under study are spatially resolved or not. The transition
from one regime to another occurs for velocities corresponding to the minimal
spatially resolved scale. The density and velocity contributions can
be separated by smoothing the data. The difference between the
VCS spectral indexes for low resolution and high resolution
depends only on the spectrum of the velocity fluctuations.

V. The transition between the low and high resolution regimes allows one
to identify the spatial scale associated with a particular turbulent velocity.
Therefore observing this transition provides a way to relate the 
velocity dispersion with the angular separation between
the lines of sight. This allows one to separate thermal and non­thermal
contribution to line broadening in clouds. 

VI. VCS allows the recovery of the underlying spectrum of turbulence
in the presence of absorption.
For the power-law spectra, the main effect of absorption amounts to
the introduction of the low frequency spatial filter for the
fluctuations along the velocity coordinate.

VII. Both in cases of negligible and important
 absorption thermal broadening of spectral lines
introduces the exponential suppression of the amplitude of
the high frequency VCS, which makes cold gas the most important
contributor to the VCS signal.
This suppression is factorized in the expressions for the VCS
and it can be corrected for by choosing the particular
exponential factor that straightens the VCS to a power law at large
$k_V$. Choosing this correction
provides yet another way of estimating the temperature of cold gas.

\acknowledgements{We thank Tom Bethel and Nicholas Hall for reading the
manuscript and providing valuable comments. We also thank Alexey Chepurnov and
two anonymous referees of the paper for their input.  Discussions with Carl
Heiles and Snezana Stanimirovic are acknowledged. AL research is supported by
by NSF grant AST 0307869 and the NSF Center for Magnetic Self Organization in
Laboratory and Astrophysical Plasmas.}

\appendix

\section{Notations}
\label{App:A}

\begin{tabular}{lp{5in}}
$\rho({\bf x})$ & 3D density field.\\
${\bf u}({\bf x})$ & 3D turbulent velocity field.\\
$\xi(r)$ & 3D density correlation function. \\
$\d(r)$ & 3D density structure function. \\
$D_z(r)$ & z-component of 3D velocity structure function. \\
$\rho_s({\bf X}_1,v_1) $ & density in PPV space .\\
$\xi_s(R,v)$ & PPV density correlation function. \\
$\d_s(R,v)$ & PPV density structure function. \\
$\tilde\xi_s(R,v)$ & PPV density fluctuations correlation function. \\
$\tilde\d_s(R,v)$ & PPV density fluctuations structure function. \\
$I({\bf X}_1,v_1)$ & emission intensity at velocity $v_1$
in the direction ${\bf X}_1$. \\
${\cal D}(v_1,v_2)$ & intensity structure function along the velocity
coordinate. \\
$P(k_v)$ & power spectrum of intensity along the velocity coordinate. \\
$\ae$ & coordinate in velocity direction rescaled to reveal P-V symmetries\\
$B({\bf X})$ & angular beam of the instrument. \\
$S$ & saturation scale of the turbulent velocities. 
Identified with the spatial extend of the turbulent cloud. \\
$D_z^{1/2}(S)$ & characteristic turbulent velocity difference
at separation $S$. Sets the extend of the spectral line along the velocity
coordinate. \\
$A_{0v}(\gamma,m)$ & $= \left\{ \begin{array}{l}
2^{\frac{3}{2}-\frac{1-\gamma}{m}}
\Gamma\left(\frac{3}{2}-\frac{1-\gamma}{m}\right)/
(1-\gamma-m/2), ~~~ m > \frac{2}{3} (1-\gamma) \\
1/(1-\gamma-3m/2), ~~~~~~~~~~~~~~~~~~~~~~~~~  m \le \frac{2}{3} (1-\gamma) \\
\end{array} \right.$. \\
$A_{R0}(\gamma,m)$&$=2 \int_0^\infty dz 
\left[z^{-\gamma-m/2} - (1+z^2)^{1-\gamma/2-m/4}/(1+m/2+z^2)\right] $. \\
$ A_{R\ae}(\theta,\gamma,m)$ & = 2 $ \int_0^{\infty} dz 
\left[z^{-\gamma-m/2} - \frac{(\sin^2\theta+z^2)^{1-\gamma/2+m/4}}{(1+m/2)\sin^2\theta+z^2}
\exp\left(-\frac{A(\gamma,m)}{2} \frac{\cos^m\theta (\sin^2\theta+z^2)^{1-m/2}}
{(1+m/2)\sin^2\theta+z^2} 
\right)\right] $. \\
$F(\gamma,m)$&$=
\left[A_{R0}(\gamma,m)/A_{0v}(\gamma,m)\right]^{m/(1-\gamma-m/2)}$. \\
$C(\gamma,m)$ & = $\sqrt{2}
\left[\Gamma(1/m)/\Gamma((1-\gamma)/m)\right]^{m/2\gamma}$ .
\end{tabular}

\section{Homogeneity of Data and Reduction of Noise}
\label{App:B}

For practical data handling one has to deal with the
actual line shape. An important issue is 
whether the statistics along the V-coordinate can be assumed homogeneous,
in which case the statistical descriptors are the functions of the
velocity separation $v=v_1-v_2$ only, ${\cal D}(v_1,v_2)={\cal D}(v)$.
To be valid this requires statistical homogeneity of the PPV density,
which, in particular, corresponds to the case of a flat mean line profile
$\langle I_{\bf X}(v) \rangle = const$ for the velocities under consideration.
More generally, one would like to subtract the mean line profile, and
assume homogeneity for the fluctuations  only.

Whether statistics is homogeneous or not
is also important to our ability to measure it with the noisy data.
In the measurements, the ensemble average $\langle\ldots\rangle$
is replaced by the volume average
over the space in which the statistics is homogeneous, i.e. does not change
with the translations in space.
Take ${\cal D}(v_1,v_2)$. How can we estimate it from the data ?
If it depends separately on $v_1$ and $v_2$ and we have only a single line of
sight, then there is no averaging available
and our estimate of the correlation
will be very noisy ${\cal D}(v_1,v_2) \approx [I(v_1)-I(v_2)]^2$.
For homogeneous statistics, however,
we can average over $v_+=(v_1+v_2)/2$,
${\cal D}(v) \propto \int dv_+ [I(v_1)-I(v_2)]^2$, beating the noise
down significantly. However, if we have measurements at several lines of sight,
we can average our estimator over ${\bf X}$ which may give us the ability
to measure even inhomogeneous ${\cal D}(v_1,v_2)$.

Statistics
of the fluctuations will then be described by the
modified structure function
\begin{equation}
\tilde {\cal D}(v) = \langle 
\left[ I(v_1) - \langle I(v_1) \rangle \right]
\left[ I(v_2) - \langle I(v_2) \rangle \right] \rangle
= {\cal D}(v_1,v_2)
- \left[\langle I(v_1) \rangle - \langle I(v_2) \rangle
\right]^2
\label{eq:tildeD}
\end{equation}
Eq.~(\ref{eq:tildeD}) indicates that 
at small separations $v$ the correction due to the mean profile 
is at least quadratic in $v$,
$\left[\langle I(v_1) \rangle - \langle I(v_2) \rangle \right]^2 \sim v^2$.
As we discuss in \S 2.5 this results in an easily separable $\delta$ function
contribution to the VCS.

\section{Statistics of density in PPV}
\label{App:C}

The random fields $ \rho({\bf x})$ and $u({\bf x})$ are assumed to be uncorrelated. The accuracy of this assumption was studied
analytically in LP00, LP04 and tested numerically in  Lazarian et al.
(2001), and Esquivel et al. (2003). As the result, this key assumption was
shown to be accurate both for the data obtained via MHD turbulence simulations
as well as for particular general type flows that were treated analytically.  
We shall also consider statistical properties of the density
distribution $\rho({\bf x})$ to be homogeneous.

Under these assumptions, the mean and the two-point correlation function of the
PPV density are given, correspondingly, by
\begin{eqnarray}
\langle \rho_s({\bf X}_1, v_1)\rangle &=&
\int_0^S dz_1 \; \langle \rho({\bf x}_1)\rangle 
\langle \phi_{v1}({\bf x}_1) \rangle =  \nonumber \\
&=&\bar\rho \int_0^S dz_1 \; \langle \phi_{v1}({\bf x}_1) \rangle ~~, \\
\langle \rho_s({\bf X}_1, v_1)
\rho_s({\bf X}_2,v_2)\rangle &=& 
\int_0^S dz_1 \int_{0}^{S} dz_2 \;
\langle \rho({\bf x}_1)\rho({\bf x}_2)\rangle \;
\langle \phi_{v1}({\bf x}_1) \phi_{v2}({\bf x}_2) \rangle = \nonumber \\
&=& \int_0^S dz_1 \int_{0}^{S} dz_2 \; \xi({\bf r})
\langle \phi_{v1}({\bf x}_1) \phi_{v2}({\bf x}_2) \rangle  ~~.
\end{eqnarray}
The computation is, thus, reduced to evaluating the ensemble average
of the products of the Maxwell functionals
$\phi_v({\bf x})$ over the Gaussian distribution of the turbulent
velocities $u$, i.e,
$\langle \phi_{v1}({\bf x}_1) \rangle =  \int du_1\phi_{v1}({\bf x}_1) P(u_1)$
and 
$\langle \phi_{v1}({\bf x}_1) \phi_{v2}({\bf x}_2) \rangle = 
\int {\mathrm d}u_1 {\mathrm d}u_2\phi_{v1}({\bf x}_1)
\phi_{v2}({\bf x}_2) P(u_1,u_2), ~u_1 = u({\bf x_1}),~u_2 = u({\bf x_2})$.
In our picture we identify the scale $S$ at
which the structure function $D_z({\bf r})$ 
that describes the turbulent field
is saturated $ D_z(\infty) \approx D_z(S)$ with the size of the emitting cloud.
The one point distribution function of the line-of-sight velocity component
is then
\begin{equation}
P(u_1) = \frac{1}{\sqrt{\pi D_z(S)}}
\exp\left[-\frac{u_1^2}{D_z({S})}\right]
\end{equation}
(the velocity is defined with respect to the rest frame of the cloud,
and is, thus, taken to have a zero mean)
while the Gaussian two point probability function is most
conveniently given in terms of the uncorrelated
variables $u=u_1-u_2$ and $u_+=(u_1+u_2)/2$ 
\begin{equation}
P(u,u_+) = \frac{1}{\pi \sqrt{2 D_z(S)-D_z({\bf r})}\sqrt{D_z({\bf r})}}
\exp\left[-\frac{u_+^2}{D_z(S)-D_z({\bf r})/2}\right]
\exp\left[-\frac{u^2}{2 D_z({\bf r})}\right]
\end{equation}

In LP04 we obtained the general results. 
Here we reproduce the results of the computation just for the case of vanishing
regular shear
 velocities $v_{gal}$ that are dealt with in this paper
\begin{eqnarray}
\langle \rho_s({\bf X}_1, v_1)\rangle & = &
\frac{\bar \rho S}{\sqrt{\pi} \left[D_z(S)+2\beta\right]^{1/2}}
\exp\left[-\frac{v_1^2}{D_z(S)+2\beta}\right]
\label{AppB:mean_cloud} \\
\langle \rho_s({\bf X}_1, v_1)
\rho_s({\bf X}_2,v_2)\rangle &=& \frac{1}{2 \pi}
\int_{-S}^S {\mathrm d}z \left(S-|z|\right)
\; \frac{\xi({\bf r})}{[D_z({\bf r})+2\beta]^{1/2}}
\exp\left[-\frac{v^2}{2 (D_z({\bf r})+2\beta)}\right] \nonumber \\ 
&\times& \frac{\sqrt{2}}{ [\beta+D_z(S)-D_z({\bf r})/2]^{1/2}}
\exp\left[-\frac{v_+^2}{\beta+D_z(S)-D_z({\bf r})/2}\right]~~,
\label{AppB:2point_cloud}
\end{eqnarray}
where we used the notation $v=v_1-v_2$, $v_+=(v_1+v_2)/2$,
${\bf r}={\bf r}_1-{\bf r}_2$, $z=z_1-z_2$, ${\bf R}={\bf X}_1-{\bf X}_2$.

The residual dependence
of the quantities in Eqs.~(\ref{AppB:mean_cloud}-\ref{AppB:2point_cloud})
on the absolute velocity $v_1$ or $v_+$ is the signature of statistical
inhomogeneity of the density in PPV space at the edges of the line
$|v_1|,|v_+| > (D_z(S)+\beta)^{1/2}$.  However, we are interested primarily in
the ${\bf X}$ and $v$ dependence of PPV correlations at separations
small compared with the extent of the cloud in PPV space.
If the measurements are done in a narrow velocity channel, then the
inhomogeneity means a different normalization of the correlation function
for channels at the edge of the line relative to central channels. 
If localization along the velocity coordinates
is not focused upon, the observed correlations can, and often are,
estimated by averaging over all central velocities $v_+$ for the fixed $v$.
Whenever the observed signal is linearly related to the density in PPV,
as in the case of the intensity in an optically thin line,
such estimation is given by the PPV density correlation averaged
along the velocity coordinate
\begin{eqnarray}
\xi_s({\bf R},v) & \approx & \frac{1}{[D_z(S)+2\beta]^{1/2}}
\int {\mathrm d}v_+ \langle \rho_s({\bf X}_1, v_1)\rho_s({\bf X}_2,v_2)\rangle
\label{eq:main_approx}
\\
&\approx&
\frac{{\bar \rho}^2 S}{ [D_z(S)+2\beta]^{1/2}}
\int_{-S}^S {\mathrm d}z \left(1-\frac{|z|}{S}\right)
\; \frac{\xi({\bf r})/\bar \rho^2}{[D_z({\bf r})+2\beta]^{1/2}}
\exp\left[-\frac{v^2}{2 (D_z({\bf r})+2\beta)}\right] \nonumber
\end{eqnarray} 
which is not sensitive to the mean line profile. In this paper we shall
not consider scales comparable to the whole extent of the line
and will use Eq.~(\ref{eq:main_approx}) as our main formula
for $v \ll D_z^{1/2}(S)$.

At vanishing temperature
\begin{equation}
\xi_s({\bf R},v) \approx 
\frac{{\bar \rho}^2 S}{ D_z^{1/2}(S)}
\int_{-S}^S {\mathrm d}z \left(1-\frac{|z|}{S}\right)
\; \frac{\xi({\bf r})/\bar \rho^2}{D_z^{1/2}({\bf r})}
\exp\left[-\frac{v^2}{2 D_z({\bf r})}\right]
\label{eq:main_T0}
\end{equation}
while for a finite temperature we can cast Eq.~(\ref{eq:main_approx}) in
the form of convolution of the zero temperature correlation function with
the thermal window
\begin{equation}
\xi_s({\bf R},v) \propto
\int_{-\infty}^\infty
\frac{dv^\prime}{(4 \pi \beta)^{1/2}} 
\exp\left[-\frac{(v-v^\prime)^2}{4 \beta}\right]
\int_{-S}^S {\mathrm d}z \left(1-\frac{|z|}{S}\right)
\; \frac{\xi( r)}{D_z({\bf r})^{1/2}}
\exp\left[-\frac{{v^\prime}^2}{2 D_z({\bf r})}\right] ~.
\label{eq:main_T}
\end{equation}

\section{1D, 2D and 3D PPV Power Spectra}
\label{App:D}

Here we present short-wave asymptotics  and useful approximations 
for the 1D, 2D and 3D PPV power spectra.
One of the main formulas of LP00, Eq.~(16) expresses the
3D PPV power spectrum through a 3D real space correlation function.
Using the notation of the present paper this result can be written as
\begin{equation}
P_s({\bf k}) \propto \int d^3 {\bf r} e^{i {\bf K R}} \xi(r)
\exp\left[-k_v^2 D_z({\bf r})\right]
\end{equation}
In comparison with LP00, 
here ${\bf k} = ({\bf K}, k_v D_z^{1/2}(S)/S)$ and in the absence
of coherent shearing motions (e.g., due to galactic rotation) the
$3D$ plane wave is reduced to the $2D$ one. 
For the negative $\gamma$ we integrate over the structure  function
$d(r)$ rather than the correlation function $\xi(r)$.

The one-dimensional spectrum along the velocity coordinate is
\begin{eqnarray}
P_1(k_v) &=& \int {\mathrm d}{\bf K} P_s({\bf K}, k_v)  \\
&\propto& \int dz \xi(z) \exp\left[-k_v^2 D_z(S)(z/S)^m\right] \nonumber \\
&\propto& \left(r_0/S\right)^\gamma 
\left[k_v D_z^{1/2}(S)\right]^{2(\gamma-1)/m}
\end{eqnarray}
The result is valid for $\gamma < 1$.
$P_1(k_v)$ coincides with $P_{nar}(k_v)$ in this paper.

The two-dimensional spectrum orthogonal to the line of sight is
\begin{eqnarray}
P_2(K) &=& \int {\mathrm d} k_v P_s({\bf K}, k_v)  \\
&\propto& \int d^3 {\bf r} e^{i {\bf K R}} \frac{\xi(r)}{r^{m/2}} 
- (regularization~for~\gamma+m/2 \le 1) \nonumber \\
\label{AppD:P2reg}
&\propto& -\frac{r_0^\gamma S^{m/2} }{D_z^{1/2}(S)} 
\int d^2 {\bf R} e^{i {\bf K R}} \int {\mathrm d} z
\left[\frac{(R^2+z^2)^{1-\gamma/2+m/4}}{(1+m/2)R^2+z^2} - \frac{1}{z^{\gamma+m/2}} \right]\\
&\propto& \left(r_0/S\right)^\gamma 
\left(K S\right)^{\gamma+m/2-3}
\label{AppD:P2asym}
\end{eqnarray}
The regularizing term that appears in Eq.~(\ref{AppD:P2reg}) is proportional to
$\delta({\bf K})$, thus not affecting the high wave number asymptotics.
Its introduction is equivalent to taking the Fourier transform
of $ d_s(R)=2(\xi_s(0,0)-\xi_s(R,0))$  rather than
$\xi_s(R,0)$ to compute the $P_2({\bf K})$. This definition is appropriate
for shallow spectra $\gamma+m/2 \le 1$. The final asymptotics is valid for
$\gamma+m/2 > -1$. For $1 < \gamma+m/2 < 3$ no regularization is necessary
and asymptotic expression~(\ref{AppD:P2asym}) arises directly.

The spectra in orthogonal PPV directions, 
$P_1(k_v)$ and $P_2(K)$ look quite different. 
Let us however introduce the scaled velocity wavenumber $k_{\ae} = S 
\left[ k_v^2 D_z(S) \right]^{1/m}$ and correspondingly transformed
PPV wavevector ${\bf k}_{\mathrm{PPV}} = ({\bf K}, k_{\ae})$. 
The 1D spectrum  in $k_{\ae}$ coordinates is obtained
from the condition of power conservation under variable change,
$P_1(k_{\ae}) {\mathrm d} k_{\ae} = P_1(k_v) {\mathrm d} k_v =
(D_z (S)/S^m)^{1/2} P_1\left(k_v[k_{\ae}]\right) k_{\ae}^{m/2-1}
{\mathrm d} k_{\ae} $. Therefore,
\begin{equation}
P_1(k_{\ae}) \propto \left(r_0/S\right)^\gamma (D_z (S)/S^m)^{1/2}
\left[k_{\ae} S\right]^{\gamma+m/2-2} ~~.
\end{equation}
With this variable change, correspondent to transformation in PPV space
from $v$ to $z_v$ according to Eq.~(\ref{eq:vzmap}),
the spectra exhibit the same scaling in both $z_v$ and position coordinates
(the difference by one is due to different dimensionality of spaces that
$P_1$ and $P_2$ are defined on).
This is the same relation that we have discussed for the structure functions in 
Section~\ref{sec:symmetry}, however the language of spectra has an advantage
that there is no issue with structure function saturation for 
$\gamma+3m/2 < 1$.

The asymptotic scalings of all PPV spectra at high wavenumbers, including
the one for 3D $P_s({\bf K},k_v)$ that we take from LP00, are
summarized  in Table~\ref{tab:1Dspk_asymp}. In this Table 
the two parts that 3D spectrum $P_s({\bf k})$ splits into (see LP00)
\begin{equation}
P_s({\bf k}) = P_v({\bf k}) + P_{\rho}({\bf k})~~~, 
\label{eq:insum}
\end{equation}
(where the part $P_v({\bf k})$ depends only on the velocity
statistics and the part $P_{\rho}({\bf k})$ has  contributions
from both velocity and density), are presented separately. 

\begin{table}[htb]
\centering
\begin{tabular}{lccc} \hline\hline\\
& \multicolumn{1}{c}{ 1D: $P_s(k_v)$} & 2D: $P_s(K)$ & 3D: $P_s(K,k_v)$  \\[2mm]
& \multicolumn{1}{c}{ $ k_v D_z^{1/2}(S) \gg 1 $} & $ KS \gg 1 $
&  $ k_v^2 D_z(S) \gg (k S)^m $ 
\\[2mm] \hline \\
$ P_{\rho}: $ & $(r_0/S)^\gamma \left[k_v D_z^{1/2}(S)\right]^{2(\gamma-1)/m} $
& $ \left(r_0/S\right)^\gamma \left[K S\right]^{\gamma+m/2-3} $
&  $ (r_0/S)^\gamma \left[k_v D_z^{1/2}(S)\right]^{-2 (3-\gamma)/m} $ \\[2mm] 
\hline \\
$ P_{v}:$ & $ \left[k_v D_z^{1/2}(S)\right]^{-2/m}$
& $\left[K S\right]^{m/2-3}$
& $\left[k_v D_z^{1/2}(S)\right]^{-6/m} $ \\[3mm]
\hline
\end{tabular}
\caption{The short-wave asymptotic behavior
of power spectra  in PPV space.}
\label{tab:1Dspk_asymp}
\end{table}

\end{document}